\documentclass[aps,prc,reprint,superscriptaddress,nofootinbib,showpacs]
{revtex4-1}

\usepackage{amsmath} 
\usepackage{bm} 
\usepackage{multirow,bigdelim}
\usepackage{color} 
\usepackage{graphicx}
\usepackage{subfigure}
\providecommand \href [0]{\begingroup \@sanitize@url \@href}

\usepackage{CJKutf8}
\begin{document}

\title{First measurement of $^{30}$S+$\alpha$ resonant elastic scattering for 
the $^{30}$S($\alpha$,\,p) reaction rate}

\author{D.~Kahl}
\email[]{daid.kahl@ed.ac.uk}
\affiliation{Center for Nuclear Study, the University of Tokyo, Wako, Saitama 
351-0198, Japan}
\affiliation{School of Physics \& Astronomy, the University of Edinburgh, 
Edinburgh EH9 3JZ, UK}

\author{H.~Yamaguchi (\begin{CJK}{UTF8}{min}山口~英斉\end{CJK})}
\affiliation{Center for Nuclear Study, the University of Tokyo, Wako, Saitama 
351-0198, Japan}

\author{S.~Kubono(\begin{CJK}{UTF8}{min}久保野~茂\end{CJK})}
\affiliation{Center for Nuclear Study, the University of Tokyo, Wako, Saitama 
351-0198, Japan}
\affiliation{RIKEN Nishina Center, Wako, Saitama 351-0198, Japan}
\affiliation{Institute of Modern Physics, Chinese Academy of Sciences, Lanzhou 
730000, China}

\author{A.~A.~Chen}
\affiliation{Department of Physics \& Astronomy, McMaster University, Hamilton, 
Ontario L8S 4M1, Canada}

\author{A.~Parikh}
\affiliation{Departament de F\'isica, Universitat Polit\`ecnica de Catalunya, 
Barcelona, Spain}

\author{D.~N.~Binh}
\thanks{Present address: 30 MeV Cyclotron Center, Tran Hung Dao Hospital, Hoan 
Kiem District, Hanoi, Vietnam.}
\affiliation{Center for Nuclear Study, the University of Tokyo, Wako, Saitama 
351-0198, Japan}

\author{J.~Chen(\begin{CJK}{UTF8}{gbsn}陈俊\end{CJK})}
\thanks{Present address: Nuclear Data Center, National Superconducting 
Cyclotron Laboratory, Michigan State University, 640 S. Shaw Ln, East Lansing, 
Michigan 48824, USA.}
\affiliation{Department of Physics \& Astronomy, McMaster University, Hamilton, 
Ontario L8S 4M1, Canada}

\author{S.~Cherubini}
\affiliation{Laboratori Nazionali del Sud-INFN, Catania, Italy}
\affiliation{Dipartimento di Fisica e Astronomia, Universit\`a di Catania, 
Catania, Italy}

\author{N.~N.~Duy}
\thanks{Present address: Department of Physics, Sungkyunkwan University, 300 
Chunchun-dong, Jangan-ku, Suwon 400-746, Republic of Korea.}
\affiliation{Department of Physics, Dong Nai University, 4 Le Quy Don, Tan Hiep 
Ward, Bien Hoa City, Dong Nai, Vietnam}
\affiliation{Institute of Physics, Vietnam Academy of Science and Technology, 
10 Dao Tan, Ba Dihn, Hanoi, Vietnam}

\author{T.~Hashimoto(\begin{CJK}{UTF8}{min}橋本~尚志\end{CJK})}
\thanks{Present address: Institute for Basic Science, Daejeon 305-811, Korea.}
\affiliation{Center for Nuclear Study, the University of Tokyo, Wako, Saitama 
351-0198, Japan}

\author{S.~Hayakawa (\begin{CJK}{UTF8}{min}早川~勢也\end{CJK})}
\affiliation{Center for Nuclear Study, the University of Tokyo, Wako, Saitama 
351-0198, Japan}

\author{N.~Iwasa(\begin{CJK}{UTF8}{min}岩佐~直仁\end{CJK})}
\affiliation{Department of Physics, Tohoku University, Sendai, Miyagi 980-8578, 
Japan}

\author{H.~S.~Jung(\begin{CJK}{UTF8}{mj}정효순\end{CJK})}
\affiliation{Department of Physics, Chung-Ang University, Korea}

\author{S.~Kato(\begin{CJK}{UTF8}{min}加藤 静吾\end{CJK})}
\affiliation{Department of Physics, Yamagata University, Japan}

\author{Y.~K.~Kwon(\begin{CJK}{UTF8}{mj}권영관\end{CJK})} 
\thanks{Present address: Institute for Basic Science, Daejeon 305-811, Korea.}
\affiliation{Department of Physics, Chung-Ang University, Korea}

\author{S.~Nishimura(\begin{CJK}{UTF8}{min}西村~俊二\end{CJK})}
\affiliation{RIKEN Nishina Center, Wako, Saitama 351-0198, Japan}

\author{S.~Ota (\begin{CJK}{UTF8}{min}大田~晋輔\end{CJK})}
\affiliation{Center for Nuclear Study, the University of Tokyo, Wako, Saitama 
351-0198, Japan}

\author{K.~Setoodehnia}
\thanks{Present address: Department of Physics, North Carolina State 
University, 2401 Stinson Dr, Raleigh, NC 27607, USA.}
\affiliation{Department of Physics \& Astronomy, McMaster University, Hamilton, 
Ontario L8S 4M1, Canada}

\author{T.~Teranishi(\begin{CJK}{UTF8}{min}寺西~高\end{CJK})} 
\affiliation{Department of Physics, Kyushu University, Fukuoka 812-8581, Japan}

\author{H.~Tokieda(\begin{CJK}{UTF8}{min}時枝~紘史\end{CJK})}
\affiliation{Center for Nuclear Study, the University of Tokyo, Wako, Saitama 
351-0198, Japan}

\author{T.~Yamada(\begin{CJK}{UTF8}{min}山田~拓\end{CJK})}
\thanks{Present address: Yokohama Semiconductor Co., Ltd, Japan.}
\affiliation{Department of Physics, Tohoku University, Sendai, Miyagi 980-8578, 
Japan}

\author{C.~C.~Yun(\begin{CJK}{UTF8}{mj}윤종철\end{CJK})}
\thanks{Present address: Institute for Basic Science, Daejeon 305-811, Korea.}
\affiliation{Department of Physics, Chung-Ang University, Korea}

\author{L.~Y.~Zhang(\begin{CJK}{UTF8}{gbsn}张立勇\end{CJK})}
\thanks{Present address: Key Laboratory of Optical Astronomy, National 
Astronomical Observatories, Chinese Academy of Sciences, Beijing 100012, China.}
\affiliation{Institute of Modern Physics, Chinese Academy of Sciences, Lanzhou 
730000, China}

\date{Submitted 11 January 2017; revised manuscript submitted 13 November 2017; published 3 January 2018}

\begin{abstract}
\begin{description}
\item[Background] Type I x-ray bursts are the most frequently observed 
thermonuclear 
explosions in the galaxy, resulting from thermonuclear runaway on the surface 
of an accreting neutron star.  
The $^{30}$S($\alpha$,\,p) reaction plays a critical role in burst models, yet 
insufficient experimental information is available to calculate a reliable, 
precise rate for this reaction.
\item[Purpose] Our measurement was conducted to search for states in $^{34}$Ar 
and determine their quantum properties.  In particular, natural-parity states 
with large $\alpha$-decay partial widths should dominate the stellar reaction 
rate. 
\item[Method] We performed the first measurement of $^{30}$S+$\alpha$ resonant 
elastic scattering up to a center-of-mass energy of 5.5~MeV using a radioactive 
ion beam.  The experiment utilized a thick gaseous active target system and 
silicon detector array in inverse kinematics.  
\item[Results] We obtained an excitation function for 
$^{30}$S($\alpha$,\,$\alpha$) near $150^{\circ}$ in the center-of-mass frame.  
The experimental data were analyzed with $R$-Matrix calculations, and we 
observed three new resonant patterns between 11.1 and 12.1~MeV, extracting 
their properties of resonance energy, widths, spin, and parity.  
\item[Conclusions] We calculated the resonant thermonuclear reaction rate of 
$^{30}$S($\alpha$,\,p) based on all available experimental data of $^{34}$Ar 
and found an upper limit about one order of magnitude larger than a rate 
determined using a statistical model.
The astrophysical impact of these two rates has been investigated through 
one-zone postprocessing type I x-ray burst calculations.  We find that our new 
upper limit for the $^{30}$S($\alpha$,\,p)$^{33}$Cl rate significantly affects 
the predicted nuclear energy generation rate during the burst.
\end{description}
\end{abstract}


\maketitle

\section{Introduction}\label{sec:intro}
Type I x-ray bursters (XRBs) are a class of astronomical objects observed to 
increase in luminosity by factors of typically tens to several hundreds 
\cite{2008ApJS..179..360G} for a short period of time (tens of seconds) with 
the photon flux peaking in the x ray and a total energy output of about 
$10^{39}$--$10^{40}$ ergs \cite{1977MNRAS.179...43L,1982ApJ...256..637A}.
The sources of such emissions repeat these outbursts typically on time scales of 
hours to days, allowing for the extensive study of the burst morphology of 
individual XRBs.
In our galaxy, over ninety such sources are presently known since their initial 
discovery some forty years ago.
XRBs are modelled very successfully as a neutron star accreting material rich 
in hydrogen and/or helium from a low-mass companion.
The accretion mechanism causes the formation of an electron-degenerate envelope 
around the neutron star, where the thin-shell instability triggers a runaway 
thermonuclear explosion at peak temperatures of $1.3-2.0$~GK 
\cite{2001PhRvL..86.3471S,2004ApJS..151...75W,2006NuPhA.777..601S,
2008ApJS..174..261F,2013PrPNP..69..225P}, which we observe as an x-ray burst.

The sharp rise of the x-ray fluence is understood to be powered by explosive 
helium burning on the neutron-deficient side of the Segr\`e chart 
\cite{1992ARNPS..42...39C,1993SSRv...62..223L,1999JPhG...25R.133W,
2006NuPhA.777..601S,2013PrPNP..69..225P}.
In a mixed hydrogen and helium shell, the explosive nucleosynthesis initially 
manifests as a series of ($\alpha$,\,p)(p,\,$\gamma$) reactions on oxygen seed 
nuclei near the proton drip line ($^{14,15}$O), called the $\alpha$p-process 
\cite{1981ApJS...45..389W}.
One such sequence in this burning pathway is
\begin{equation}
\begin{split}
3\alpha\rightarrow~^{12}{\rm C}({\rm p},\gamma)^{13}{\rm N}({\rm 
p},\gamma)^{14}{\rm O}(\alpha,{\rm p})^{17}{\rm F}({\rm p},\gamma)\\
^{18}{\rm Ne}(\alpha,{\rm p})^{21}{\rm Na}({\rm p},\gamma)^{22}{\rm 
Mg}(\alpha,{\rm p})^{25}{\rm Al},
\end{split}
\end{equation}
which continues as
\begin{equation}
\begin{split}
^{25}{\rm Al}({\rm p},\gamma)^{26}{\rm Si}(\alpha,{\rm p})^{29}{\rm P}({\rm 
p},\gamma)\boldsymbol{^{30}{\rm S}(\alpha,{\rm p})^{33}{\rm Cl}}({\rm 
p},\gamma)\\
^{34}{\rm Ar}(\alpha,{\rm p})^{37}{\rm K}({\rm p},\gamma)^{38}{\rm 
Ca}(\alpha,{\rm p})^{41}{\rm Sc}.
\end{split}
\end{equation}
In this sequence, the ($\alpha$,\,p) reactions proceed through 
$T_{z}=\frac{N-Z}{2}=-1$ compound nuclei.
The $\alpha$p-process gives way to the rapid proton-capture process ({\it 
rp} process) near the $Z\approx20$ region owing to the ever increasing Coulomb 
barrier and decreasing ($\alpha$,\,p) $Q$ values.
Aside from the two protons consumed in the nuclear trajectory from $^{12}$C to 
$^{14}$O, the $\alpha$p process is schematically pure helium burning (since the 
abundance of hydrogen is constant), and it does not include any $\beta^{+}$ 
decays which tend to hamper the energy generation rate in explosive 
nucleosynthesis.

While a plethora of nuclear processes tend to take place in a given regime of 
stellar nucleosynthesis, typically the precise rates of only a handful of these 
processes influence the predicted nature and magnitude of actual astrophysical 
observables. 
It is these specific nuclear quantities which should be well constrained by 
laboratory experimentation.
This general picture is confirmed in XRBs, where the nuclear reaction network 
includes hundreds of species and thousands of nuclear transmutations.
Studies have shown that it is only a small subset of these nuclear 
transmutations which need to be known precisely, as they make a predominant 
contribution to the nuclear trajectory to higher mass and energy generation 
\cite{2008ApJS..178..110P}, at least for the examined models.

The $^{30}$S($\alpha$,\,p) reaction is identified as one such important 
reaction, contributing more than 5\% to the total energy generation 
\cite{2008ApJS..178..110P}, influencing the elemental abundances in the burst 
ashes 
\cite{2008ApJS..178..110P} relevant to compositional inertia 
(see, e.g., \cite{1980ApJ...241..358T} for a description of this 
phenomenon), moving material away from the $^{30}$S waiting 
point \cite{1999ApJ...524..434I}, and possibly accounting for double peaked 
XRBs \cite{2004ApJ...608L..61F}.
A recent study found the $^{30}$S($\alpha$,\,p) reaction sensitivity in XRBs 
among the top four in a single zone model \cite{2016ApJ...830...55C}, as well 
as having a prominent (but unquantified) impact on the burst light curve in a 
multizone model.

A firmer understanding of the input nuclear physics for XRB models will allow 
for more reliable comparison with observations to constrain neutron star binary 
system properties, such as accretion rate and metallicity, as well as the 
neutron star radius itself \cite{2007ApJ...671L.141H, 2004ApJ...601..466G, 
2006Natur.441.1115O, 2012ApJ...748....5O, 2010ApJ...712..964G, 
2012ApJ...749...69Z}.

The ($\alpha$,\,p) reactions occurring on lower mass nuclei such as $^{14}$O 
and $^{18}$Ne have been measured directly \cite{2004NuPhA.746..113N, 
2007PhRvC..76b1603F, 2015PhRvC..92c5801K, 1999PhRvC..59.3402B, 
2002PhRvC..66e5802G}, and the properties of resonances in the compound nuclei 
$^{18}$Ne and $^{22}$Mg 
have been the subject of a number of indirect studies (see, e.g., Refs. 
\cite{2014PhRvC..90b5803H, 2014PhRvC..89a5804Z} and references therein). 
In spite of these extensive works, those cross sections still remain quite 
uncertain.
Unfortunately, the situation is much more dire in the case of the 
($\alpha$,\,p) reactions induced on higher mass targets such as $^{30}$S.
The only experimental information on the structure of $^{34}$Ar above the 
$\alpha$ threshold and the $^{30}$S($\alpha$,\,p) stellar reaction rate is 
limited to a preliminary report on a transfer reaction study of the compound 
nucleus $^{34}$Ar at high excitation energy \cite{2009AIPC.1090..288O} and a 
time-reversal study \cite{2011PhRvC..84d5802D}. 
The present work is the first experimental investigation using the entrance 
channel $^{30}$S+$\alpha$.

\section{Experiment}
\begin{figure*}
\includegraphics[angle=0, scale=.13, clip=true, trim=0 0 0 0]{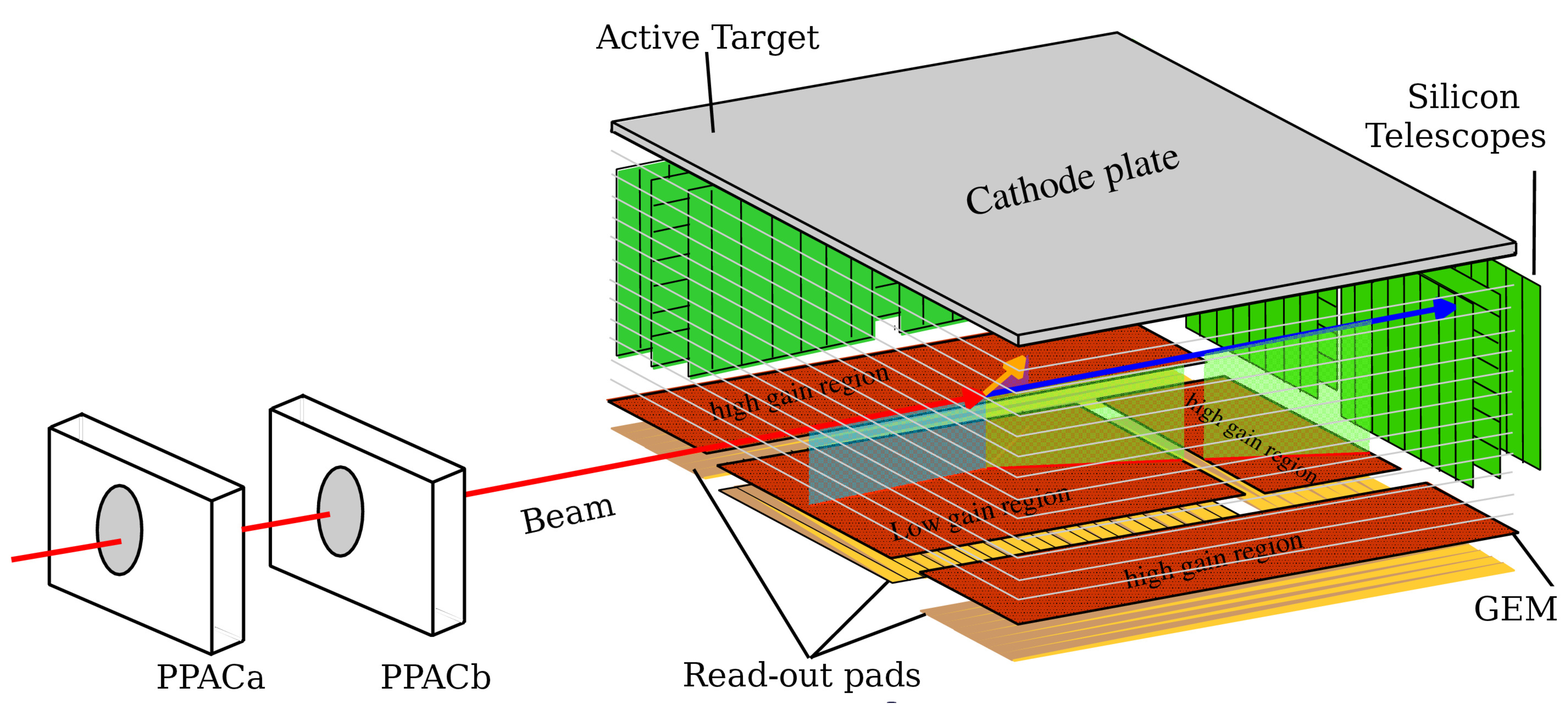}
\caption
{
Schematic of the experimental setup, consisting of two parallel plate avalanche counters (PPACs), 
the active target, and silicon telescope arrays.  Note that between PPACb and 
the active target, the beam impinges on an entrance window, which retains the 
active target fill gas.  The beam is tracked in the central low-gain region 
(``active target region'', 20 cm), surrounded on three sides by high-gain 
regions and silicon telescopes to measure outgoing light ions (right side 
telescope not depicted).  Beneath each gas electron multiplier (GEM) is a readout pattern, separated into 
4~mm thick backgammon pads.  $\Delta E$ is simply proportional to the charge 
collected by each pad.  The coordinate system is one where the beam axis 
defines positive $Z$, the rest following left-handed conventions.  $Z$ and $X$ 
positions are determined by the pad number and comparing charge collection on 
either side of the backgammon, respectively.  The $Y$ position is determined by 
the electron drift time.
}
\label{kahl:fig1}
\end{figure*}

We performed the first measurement of $\alpha$ resonant elastic scattering on a 
$^{30}$S radioactive isotope beam (RIB) using a thick target in inverse 
kinematics \cite{1990SvJNP..52..408}. 
The experiment was carried out at the CNS Radioactive Ion Beam separator (CRIB) 
\cite{2002EPJA...13..217K, 2005NIMPA.539...74Y}, owned and operated by the 
Center for Nuclear Study (CNS), the University of Tokyo, and located in the 
RIKEN Nishina Center.
The CRIB facility has been a workhorse for measurements of elastic scattering 
of primarily astrophysical interest 
\cite{2003PhLB..556...27T,2007PhLB..650..129T,2009PhLB..672..230Y,
2007PhRvC..76e5802H,2008EPJA...36....1H,2009PhRvC..80a5801H,2010.Aram.14O,
2011PhRvC..83c4306Y,
2012PhRvC..85d5802J,2012PhRvC..85a5805C,2013PhRvC..87c4303Y,2013PhRvC..88c5801J,
2013PhRvC..88a2801H,2012PhRvC..85a5805C,2012PhRvC..85d5802J,
2014PhRvC..90c5805J}, schematically using similar techniques to the 
one adopted in the present study.

The $^{30}$S RIB was produced inflight using the 
$^{3}$He($^{28}$Si,\,$^{30}$S)n transfer reaction.
A $^{28}$Si$^{9+}$ primary beam was extracted from an electron cycltron resonance ion source and 
accelerated to 7.3~MeV/u by the RIKEN AVF cyclotron ($K\approx70$) with a 
typical intensity of 80 pnA.
We impinged the $^{28}$Si beam on the production target located at the entrance 
focal plane to CRIB, comprised of a windowed, cryogenic gas cell 
\cite{2008NIMPA.589..150Y}.
$^{3}$He gas at 400~Torr was cooled to an effective temperature of 90 K with 
LN$_{2}$; the gas was confined by 2.5-$\mu$m Havar windows in a cylindrical 
chamber with a length of 80~mm and a diameter of 20~mm, yielding a $^{3}$He 
target thickness of approximately 1.7~mg~cm$^{-2}$.
As the fully stripped species $^{30}$S$^{16+}$ is the easiest to separate and 
distinguish from the primary beam, we used Be (2.5~$\mu$m) and C 
(300~$\mu$g~cm$^{-2}$) stripper foils immediately after the production target; 
when the Be (C) stripper foil was new, the $^{30}$S$^{16+}$ purity was 88\% 
(67\%), but decreased dramatically within hours as the beam degraded the foils. 
Studies are ongoing to investigate the effects of stripper foil degradation on 
beam purity and intensity.
The resulting cocktail beam was separated by a double achromatic system (set to 
$\frac{\Delta p}{p}=1.875\%$ with slits at the dispersive focal plane) and 
further purified with a Wien (velocity) filter.
The $^{30}$S RI beam arrived on target with typical purity of 28\% and an 
intensity of $8\times10^{3}$ pps, successfully injecting $1.6\times10^{9}$ 
$^{30}$S ions during the main measurement over two days.
\begin{figure*}
\includegraphics[angle=0, scale=.5, clip=true, trim=0 40 0 
0]{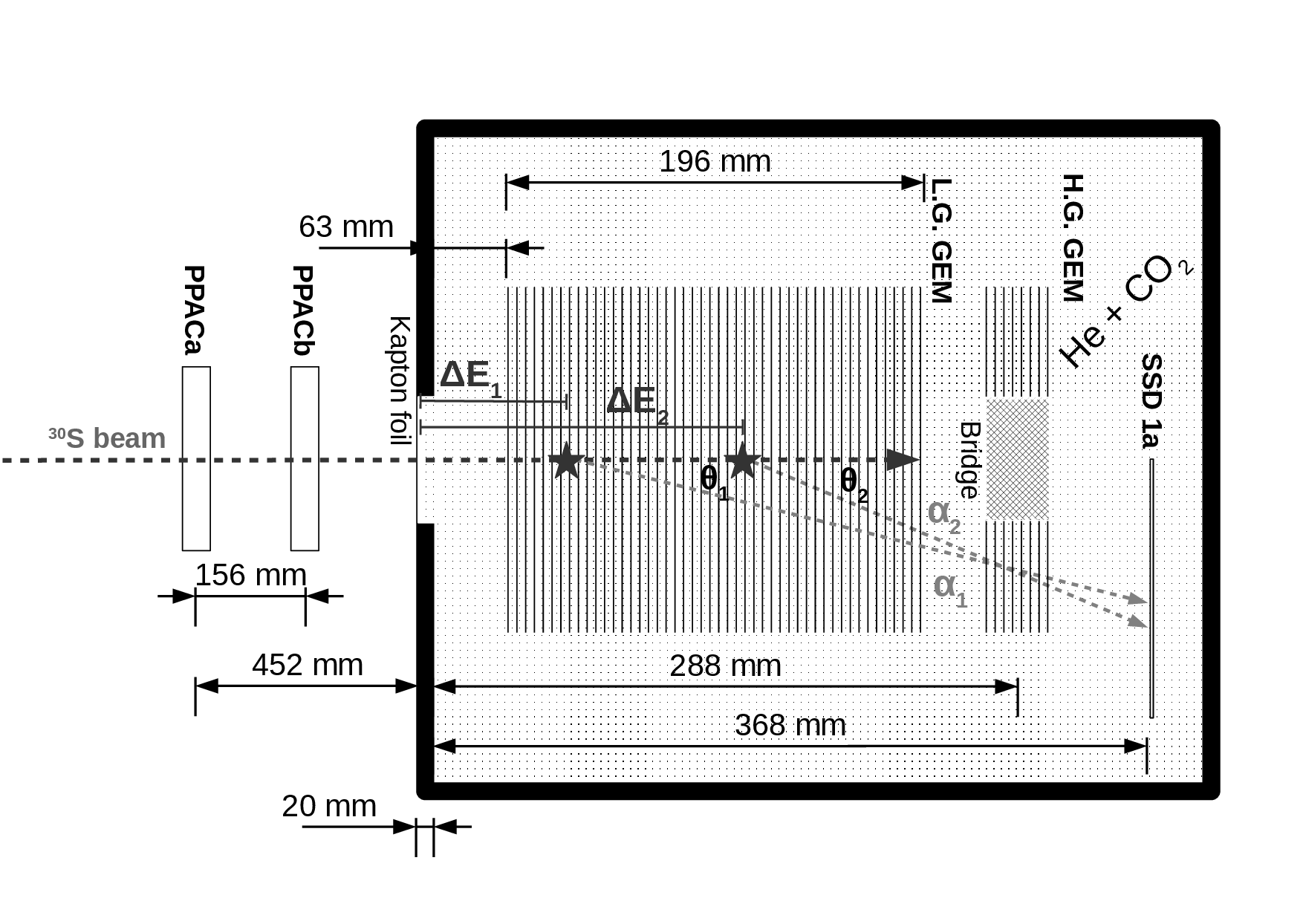}
\caption
{
Top-down cartoon of the experimental setup and the detectors used in the data 
analysis (not to scale 
-- PPACs in particular are much further from the active target chamber than 
depicted). 
 The differences between a higher energy scattering (denoted `$\alpha_{1}$') 
and a lower energy scattering (`$\alpha_{2}$') are shown.
}
\label{kahl:fig2}
\end{figure*}
\begin{figure}
 \includegraphics[scale=0.4, trim= 0 0 0 0]{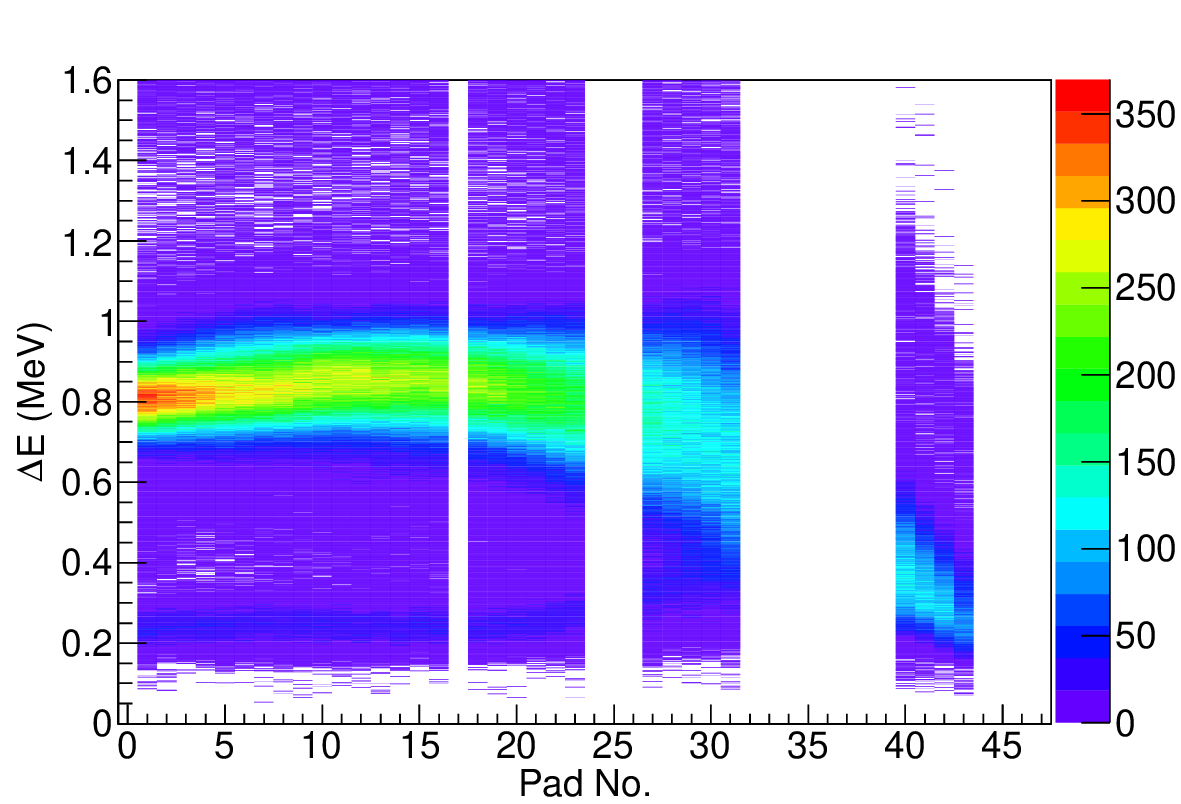}
\caption[Calibrated $^{30}$S Bragg curve]
{
Calibrated Bragg curve of the unscattered $^{30}$S beam over the 
low-gain region of the active target.  The depth of each pad is 4~mm is with 
the beam 
penetration depth correlated with the pad number.  Data from 
several pads are not shown for a variety of reasons; in general it was either 
because the electronics did not record a signal, or the energy deposit was 
arbitrarily lower than expected.
}
\label{fig:bragg}
\end{figure}
\begin{figure}
\centering
 \includegraphics[scale=0.4, trim= 0 0 0 0]{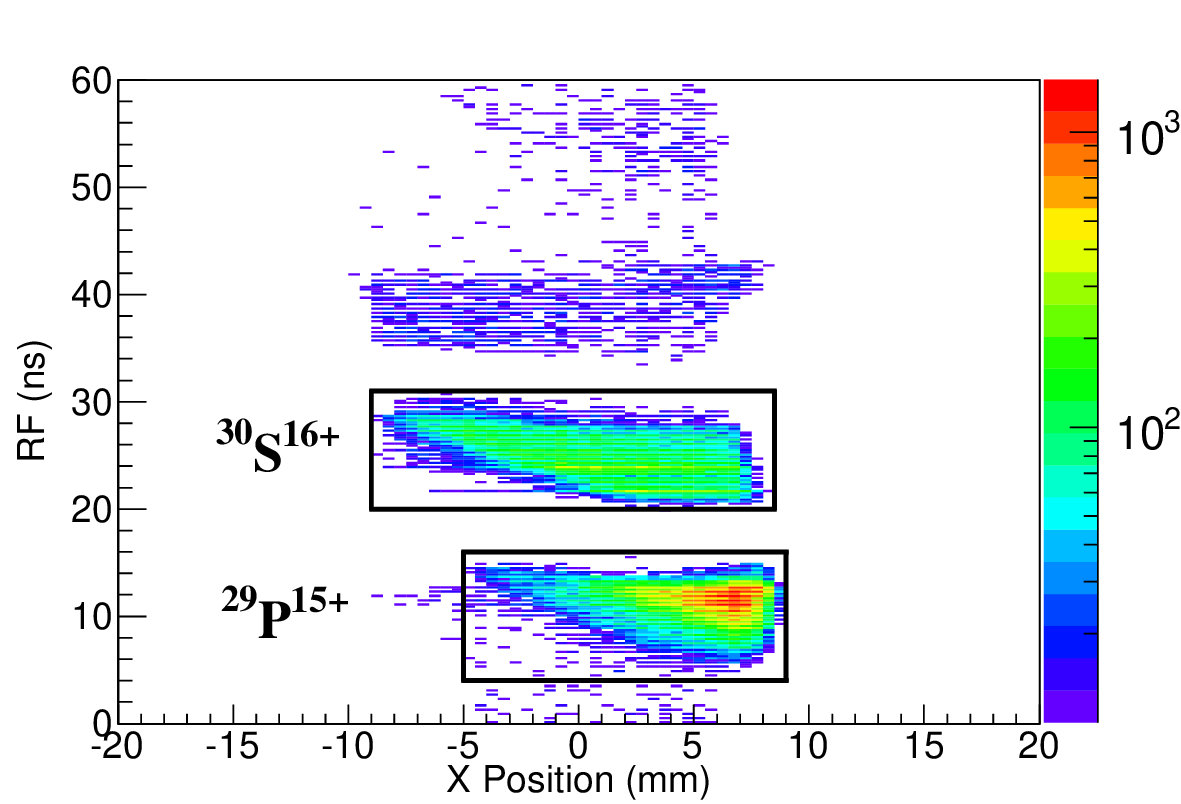} 
\caption
{Timing from $rf$ versus PPACa $X$ position for the unscattered beam, 
showing gates for $^{30}$S and $^{29}$P.  The $rf$ signal is recorded with 
PPACa as the start and the cyclotron radio-frequency signal as the stop, and 
thus it represents a relative flight time between ions in the cocktail beam.}
\label{beam-pid2}
\end{figure}

\begin{figure}
\centering
 \includegraphics[scale=0.4, trim= 0 0 0 0]{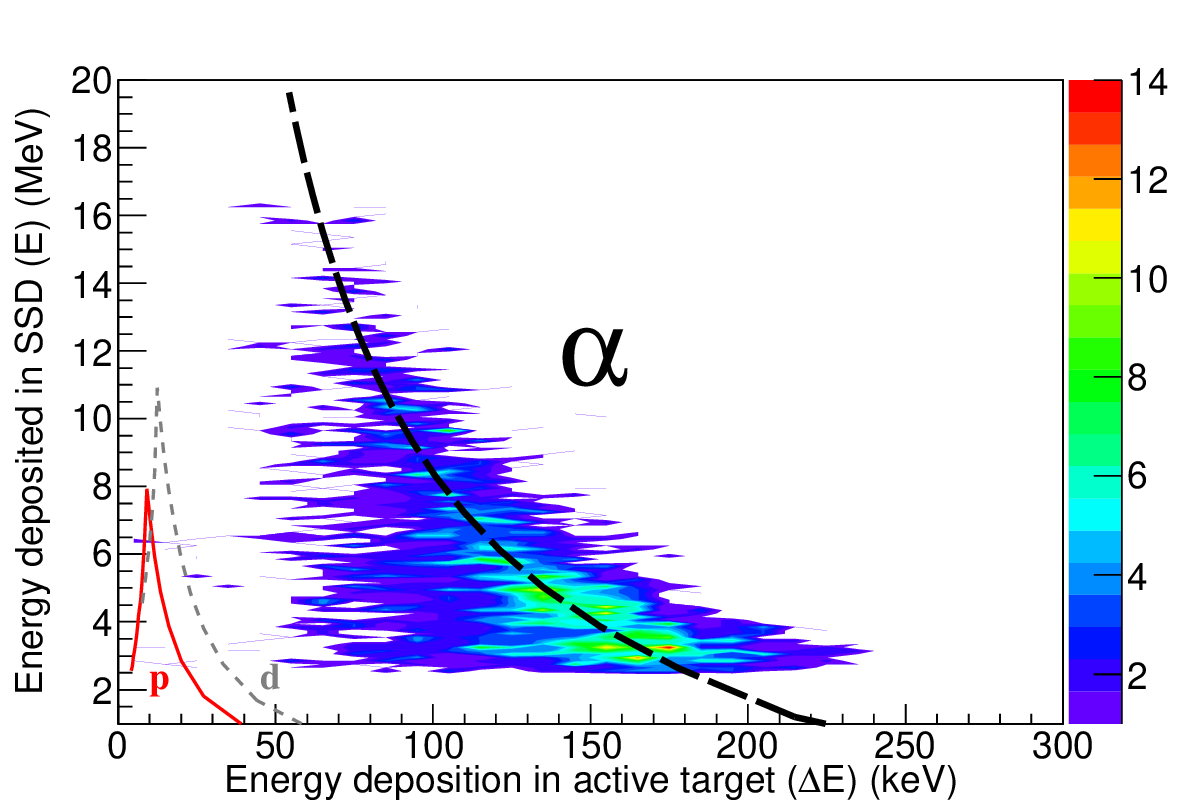} 
\caption
{$\Delta E$--$E$ plot for light ion particle identification 
during the $^{30}$S+$\alpha$ scattering measurement.  The long-dashed black 
line, short-dashed 
grey line, and solid (red) line show calculations for $\alpha$-particles, 
deuterons, and protons, respectively, using the experimental conditions.
}
\label{de-e}
\end{figure}

The setup at the experimental focal plane, shown in Figs.~\ref{kahl:fig1} \& 
\ref{kahl:fig2}, consisted of two beamline monitors, an active target system 
(see below), and an array of silicon strip detectors (SSDs).
The beamline monitors were parallel plate avalanche counters (PPACs, enumerated 
`a' and `b', respectively) \cite{2001NIMPA.470..562K}, which served to track 
the beam ions event by event and to produce the trigger signal for the
data acquisition system (DAQ).
The DAQ triggered when both PPACs fired in coincidence with an SSD to obtain 
the physics of interest; the DAQ also triggered in a downscaled mode for 
$\frac{1}{n}$ PPAC coincidences ($n=2.2 \times 10^{4}$) regardless of the SSD 
signals.
This trigger setup is standard in CRIB experiments, and it allows for
event-by-event analysis of scattering events as well as 
simultaneous diagnosis of the RIB
for systematic behavior while keeping a total trigger rate $< 1$~kHz.
Due to a DAQ error, we could only fully analyze the data from the forward 
right-side SSD.

During the RIB production, the efficiency of each PPAC was determined to be 
quite 
high ($>\!\!99$\%) considering the number of events recorded by each detector;
however, at the start of the scattering measurement PPACb was damaged by 
several 
discharges, and its efficiency became somewhat lower ($\sim90$\%) for most of 
the experiment.
Each SSD was 0.5~mm thick and had an active area of $91 \times 91$~mm$^{2}$, eight 
strips on one side, and a single pad on the reverse. 
A scattering chamber filled with about $\frac{1}{4}$ atm 90\% $^{\rm nat}$He + 
10\% 
CO$_{2}$ gas mixture housed both the active target system and the 
SSDs.\footnote{Gas mixture percentages are quoted by volume.}
The He + CO$_{2}$ gas pressure was monitored continuously throughout the 
scattering measurement and managed by a dedicated system; we set the gas flow 
controller to circulate fresh gas into the chamber at 20 standard liters per 
minute with the evacuation rate regulated to keep a constant gas pressure of 
$194.2\pm0.5$~Torr during the entire measurement.
The gas-filled chamber was sealed off from the beamline vacuum with a 7.4-$\mu$m 
Kapton foil; the entrance window was 40~mm in diameter.

An active target is a device where a material serves simultaneously as a target 
and part of a detector, in principle allowing one to perform direct 
measurements at a beam interaction position. 
The readout section of our active target is an etched copper plate placed under 
the field cage, opposite to the cathode top plate, so that electrons created in 
the electric field of the cage by ionizing radiation drift towards it.
The readout pads are separated into four sections: one for detecting the beam 
or heavy recoils and three for detecting outgoing light ions. 
Forty-eight pads comprise the beam readout section, while the regions for 
detecting light ions are comprised of eight rectangular pads each.
The pads are 3.5~mm in depth, surrounded by 0.25~mm of insulation on all sides 
(making 0.5~mm of insulation between each pad).
Each pad is also bisected diagonally into two congruent right triangles, so 
that the collected charge can be read out from two opposing sides (backgammon 
pads).
The section for detecting the beam ions is the largest and located at the 
center, slightly shifted towards the beam upstream direction after installation 
in the scattering chamber.
The regions for detecting light ions surround the beam section on the left, 
right, and downstream sides.
Gas electron multiplier (GEM) foils were used to set different effective gains 
over the beam and light-ion regions.
Over the center of the downstream high-gain GEMs was a bridge to prevent the 
unscattered beam ions from saturating the light ion signals.

We quantified the measurement capabilities of the active target using both 
online and offline measurements.
For the low-gain region, we compared the position of $^{30}$S ions derived from 
the active target to those determined by extrapolation of the PPAC data.
For the high-gain region, we analyzed the aggregate track width of radiation 
emitted from a standard $\alpha$ source in a fixed position as measured by the 
active target; the tracks were software gated to be in coincidence with a 
geometrically-central SSD strip.
By assuming a standard CRIB PPAC resolution of $\sim\!\!1~{\rm mm}$ based on 
our experience and the known finite strip size 
of the SSD, we varied the active target resolution in a Monte Carlo simulation 
until the calculations agreed with the data.
The performance of the active target depended on the type of measurement, 
quoted below at $1\sigma$.
The $Y$ position, determined by the electron drift time, was the most precise 
being $\leq 0.5$~mm; the high precision of the drift-time measurement enabled 
us to confirm the PPAC resolution, which was found to be 0.9~mm.
The $X$-position resolution, determined by charge division in the backgammon 
pads, was 3~mm.

In the present work, the typical $^{30}$S scattering laboratory angle and 
change in energy loss was difficult to reliably distinguish from the 
unscattered beam given the above resolution for the low-gain GEM in $X$.

Considering the close spacing of the high-gain GEM data and their relatively 
large distance from typical scattering locations, extrapolating the vector of 
an $\alpha$ particle's track 
results in a large uncertainty.  
Instead, we found that averaging the pad $X$ and $Y$ data over the center (in 
$Z$) of the high-gain GEM reduced the uncertainty and was sufficient for our 
purposes.

The $^{30}$S energy on target was measured to be $48.4\pm2.0$~MeV.
The stopping power for $^{30}$S in the He+CO$_{2}$ gas mixture was determined 
by both a direct measurement of the beam energy at five target pressures and by 
a comparison of the shape of the Bragg curve and stopping position of the 
unscattered ions in the active target as shown in Fig.~\ref{fig:bragg}. 
Excellent agreement was found between the measurements and the prediction using 
Ziegler's method; the maximum difference between the measured and calculated 
$^{30}$S residual energy was 700 keV or less than 4\%.
The energy loss and the Bragg curve of the contaminant $^{29}$P were also 
reproduced using an identical approach, giving us confidence in our treatment 
of the energy loss in the PPACs, entrance window, and He+CO$_{2}$ gas mixture.
The event-by-event particle identification of the cocktail beam is shown in 
Fig.~\ref{beam-pid2}.

We confirmed the energy loss of $\alpha$ particles using a standard triple 
$\alpha$ source and an $\alpha$ beam created by CRIB, checking that both their 
Bragg curves and residual energies agreed with the calculations.
A calibrated $\Delta E$--$E$ spectrum from the $^{30}$S+$\alpha$ scattering 
measurement is shown in 
Fig.~\ref{de-e}; the figure shows clearly that the measured locus is consistent 
with the theoretical trend for $\alpha$-particles.
The dynamic range of the high-gain GEMs was optimized to be 10--100~keV 
corresponding to the energy deposit of $\alpha$ particles, which would always 
be stopped in the first SSD layer unlike high energy protons.
As protons with enough energy to reach an SSD deposit $<5$~keV per pad, they 
could not be detected by the active target system.

\section{Analysis}
\subsection{Determination of cross section}
\begin{figure}
    \includegraphics[angle=0, trim=0 0 0 0,scale=.4]{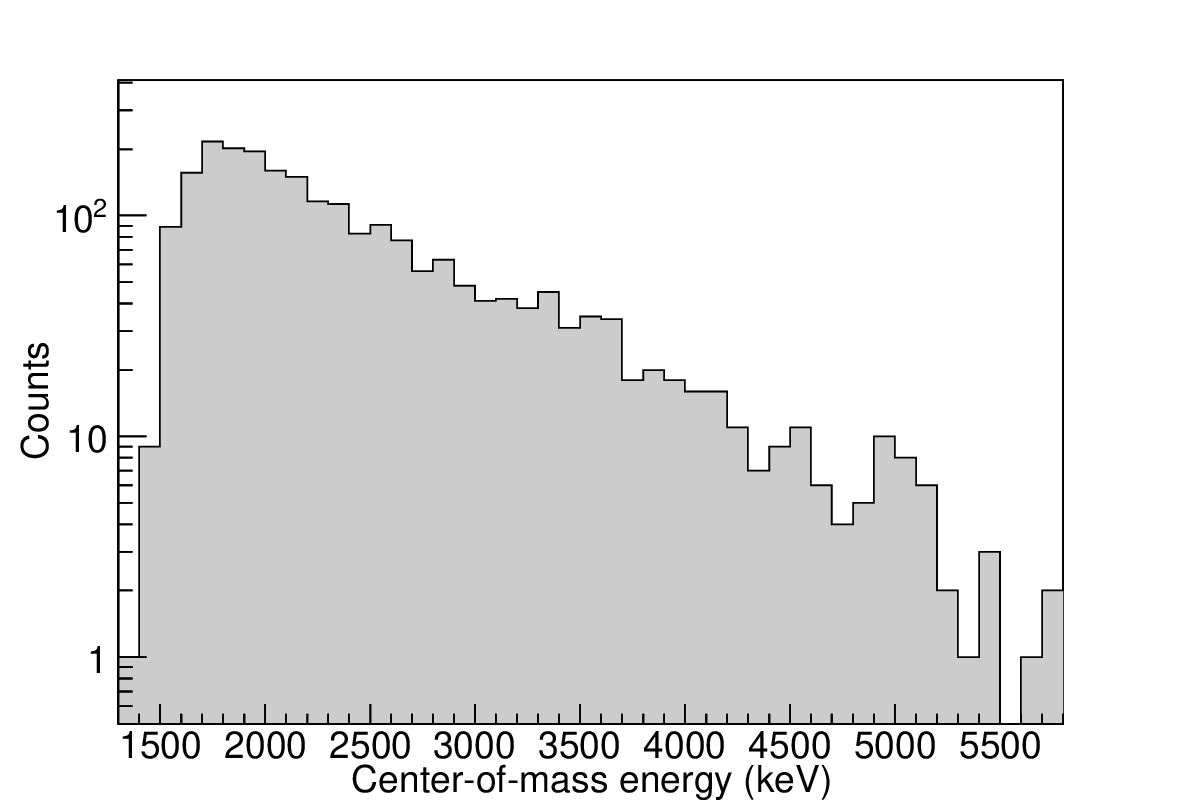}
      \caption[Adopted $R$-Matrix fit]{
      Energy spectrum of scattered $\alpha$ particles gating on the $^{30}$S 
RIB determined by the kinematic solution.  As the high-gain GEM and SSD must 
both be hit for an event to register, it is only a portion of the total events 
which are analyzed.  Hints of some resonant structure can be seen around 4500 
and 5000~keV.  The data cut off at low energy as the scattered $\alpha$ 
particles do not have enough energy to reach the SSD and are instead stopped in 
the gas.
      }
      \label{fig:alpha_spec}
      \end{figure}

We measured (1) the residual energy of $\alpha$ particles using an SSD, (2)
the beam trajectory using the two PPACs, and (3) the recoiling $\alpha$ particle
position using the high-gain portion of the active target.
These three pieces of information are sufficient to determine the 
center-of-mass energy $E_{\rm 
c.m.}$ for elastic scattering, defined as
\begin{equation}\label{eq:kinematics}
E_{\rm c.m.}=\frac{M+m}{4M\cos^{2}\vartheta_{\rm lab}}E_{\alpha},
\end{equation}
where $M$ and $m$ are the masses of $^{30}$S and $^{4}$He, respectively, 
$\vartheta_{\rm lab}$ is the laboratory scattering angle, and $E_{\alpha}$ is 
the laboratory energy of the scattered $\alpha$ particle.
Using the experimentally verified stopping power of the He+CO$_{2}$ gas for 
$^{30}$S ions and $\alpha$ particles, we numerically solved the above kinematic 
equation event-by-event.
We selected test points along the extrapolated $^{30}$S ion 
trajectory in 1~mm steps, calculating $E_{\rm c.m.}$ according to the beam's 
energy loss up to each point.
Each tested scattering depth fixes the value of $\vartheta_{\rm lab}$ by the 
geometric measurements of the PPACs and high-gain GEM.
Finally, the initial $\alpha$ energy was determined with Eq.~(\ref{eq:kinematics}) 
and its residual energy was calculated considering its energy loss along said 
path, which we then compared with the energy recorded by the SSD.
The algorithm continued until the $^{30}$S ion came to rest with no solution 
being found or when the calculated and measured residual $\alpha$ particle 
energy disagreed by less than 100~keV, which we define as a true scattering 
event at that $E_{\rm c.m.}$.
In the case that more than one test point satisfied these conditions, we select 
the $E_{\rm c.m.}$ with the smallest disagreement between the measured and 
calculated residual $\alpha$ energy.
The resulting $\alpha$ spectrum is shown in Fig.~\ref{fig:alpha_spec}.

The differential cross section was then calculated using
\begin{equation}\label{eq:dsigma-domega}
\frac{d\sigma}{d\Omega} = \frac{Y_{\alpha}S(E_{\rm b})}{I_{\rm b}n\Delta E 
\Delta\Omega_{\rm c.m.}} \frac{m}{M+m},
\end{equation}
where $Y_{\alpha}$ is the yield of $\alpha$ particles at each energy bin, 
$S(E_{\rm b})$ is the stopping power of $^{30}$S in He+CO$_{2}$, $I_{\rm b}$ is 
the number of $^{30}$S beam ions injected into the target, $n$ is the number 
density of $^{4}$He atoms, $\Delta E$ is the energy bin size (100~keV), and 
$\Delta \Omega_{\rm c.m.}$ is the center-of-mass solid angle at each energy bin.
The number of beam ions injected into the target $I_{\rm b}$ was defined as the 
coincidence between the two PPACs, recorded as a scaler during the run, 
multiplied by the average $^{30}$S purity, which includes a cut for the active 
target 
entrance window for successful injection into the target. 
The lower efficiency of PPACb cancels out in the deduction of the cross section 
because
we demanded beam particle detection with the PPACs for counting both the number 
of 
scattering events and injected beam ions.
Since the scattering could take place over a range of target depths, we 
calculated the solid angle $\Omega_{\rm c.m.}$ from the vantage point of each 
actual scattering event and fit the trend with an empirical function.
The yield of $\alpha$ particles $Y_{\alpha}$ was scaled universally by a factor 
of $2.0\pm0.1$ to match the calculated magnitude of Rutherford scattering at 
lower 
energies; a similar deficiency was observed in the number of $\alpha$ particles 
(produced in the cocktail beam by CRIB) detected by the high-gain GEM compared 
to the SSD in a low-statistics test run giving a scaling factor 1.8$\pm$0.3 
(see Secs. II~B and III~D below and Ref. \cite{kahlphd} for 
further details).
A series of measurements are already planned to further investigate and 
constrain
this scaling factor.
The resulting excitation function is shown in Fig.~\ref{rmatrix-good}(a). 
\subsection{Sources of background}
Detected $\alpha$ particles might originate from a source other than elastic 
scattering of $^{30}$S with the helium nuclei in the target gas.
We applied software gates to the PPAC data event-by-event to ensure the 
incident beam ions were consistent with the properties of $^{30}$S, which 
removed contributions to the $\alpha$ spectrum induced by other heavy ion 
species within the cocktail beam.

One might imagine various reactions with the PPACs, Kapton window 
(stoichiometry C$_{22}$H$_{10}$N$_{2}$O$_{5}$), or the CO$_{2}$ used as a 
quenching gas in the active target.  The standard PPACs used at CRIB are each 
filled with 9~Torr C$_{3}$F$_{8}$ over a length of $\approx 35$ mm ($\approx 
0.3~{\rm mg}~{\rm cm}^{-2}$) confined with 2~$\mu$m aluminized Mylar windows 
(H$_{8}$C$_{10}$O$_{4}$) and interspaced with a further three 1.5-$\mu$m 
similar foils (8.5~$\mu$m in total).

The $^{30}$S beam profile on PPACa does not have a line of sight to the 
high-gain GEM owing to the active target entrance window combined with the 
bridge over the downstream high-gain GEM, although the edge of the $^{30}$S 
profile on PPACb does have 
such a line of sight.  Thus, we can geometrically rule out PPACa (but not 
PPACb) as a source of background $\alpha$ particles.

Although the CNO-group elements require some consideration, we can immediately 
rule out hydrogen as a background source of $\alpha$ particles, because the 
$^{30}$S(p,\,$\alpha$) reaction $Q$ value is $-8.47$~MeV, and the $^{30}$S+p 
system $E_{\rm c.m.}<4$~MeV anywhere after the dispersive focal plane.

As for the entrance window and quenching gas, the Coulomb barriers for 
$^{30}$S+$^{12}$C, $^{30}$S+$^{14}$N, and $^{30}$S+$^{16}$O are 24.4, 28.0, and 
31.3~MeV, respectively.  
The $^{30}$S beam energy impinging on the Kapton window is about 2.34~MeV/u, 
yielding $E_{\rm c.m.}=20.0, 22.3, 24.3$~MeV for nuclear interactions with 
$^{12}$C, $^{14}$N, and $^{16}$O, respectively.
As for the incident $^{30}$S beam energy impinging on the He+CO$_{2}$ gas, it 
is about 1.62~MeV/u, which yields $E_{\rm c.m.}=13.9, 16.9$~MeV for nuclear 
interactions with $^{12}$C and $^{16}$O, respectively.
Considering that the center-of-mass energies are always below the respective 
Coulomb barriers for the entrance window and quenching gas, this implies that 
the heavy-ion fusion cross sections should be many orders of magnitude lower 
than that of $\alpha$ elastic scattering.

Although we are not aware of any experimental data concerning $^{30}$S-induced 
heavy-ion reactions, the fusion study with $^{12}$C and $^{16}$O on the stable 
isotopes $^{28,29,30}$Si by Jordan {\em et al.} \cite{1979PhLB...87...38J} is 
analogous if we accept isospin symmetry.  
Their center-of-mass energies broadly overlap with ours sufficiently to make a 
germane comparison.  
In that work, the authors see smooth behavior of the excitation functions 
except in the case of $^{12}$C+$^{28}$Si, where they report oscillatory 
behavior 
between $21 \sim 26$ MeV in $^{12}$C bombarding energies, which would 
correspond to maximum $E_{\alpha}$ in the range of $5.5 \sim 7$ MeV from the 
$^{28}$Si($^{12}$C,\,$^{4}$He) reaction.
$E_{\alpha}$ corresponding to the lowest region of interest in our study of 
$^{30}$S+$\alpha$ is around 15 MeV ($E_{\rm c.m.} \approx 4$ MeV).
Thus, the only structure seen by Jordan {\it et al.} is quite far away from the 
structure we report here.

As relevant to the present study, 
Jordan {\em et al.} importantly find smooth cross sections for $^{12}$C and 
$^{16}$O with 
$^{30}$Si, the mirror nucleus of $^{30}$S.  
Such behavior implies that a background source of $\alpha$ particles in the 
present work induced by CNO-group elements should have a relatively flat energy 
distribution.  
Figure~\ref{rmatrix-good}(a) shows that our observed resonant structure is 
manifested as destructive interference with pure Rutherford scattering.  
It means that any unaccounted for background of $\alpha$ particles arising from 
the $^{30}$S beam interacting with CNO-group elements would tend to decrease 
our observed resonance dips and thus our deduced partial widths $\Gamma_{i}$ 
could be modestly smaller than the true values.
If we consider the relative differences in the maximum ($\approx$~60~mb/sr) and 
minimum ($\approx$~5~mb/sr) differential cross sections around 5~MeV 
center-of-mass energy, then a smooth background cannot comprise more than 8\% 
of the measurement in the region of interest.
This uncertainty turns out to be smaller than the statistical error and as such 
can be reasonably neglected.

\begin{figure}
 \includegraphics[scale=0.4, trim= 0 0 0 0]{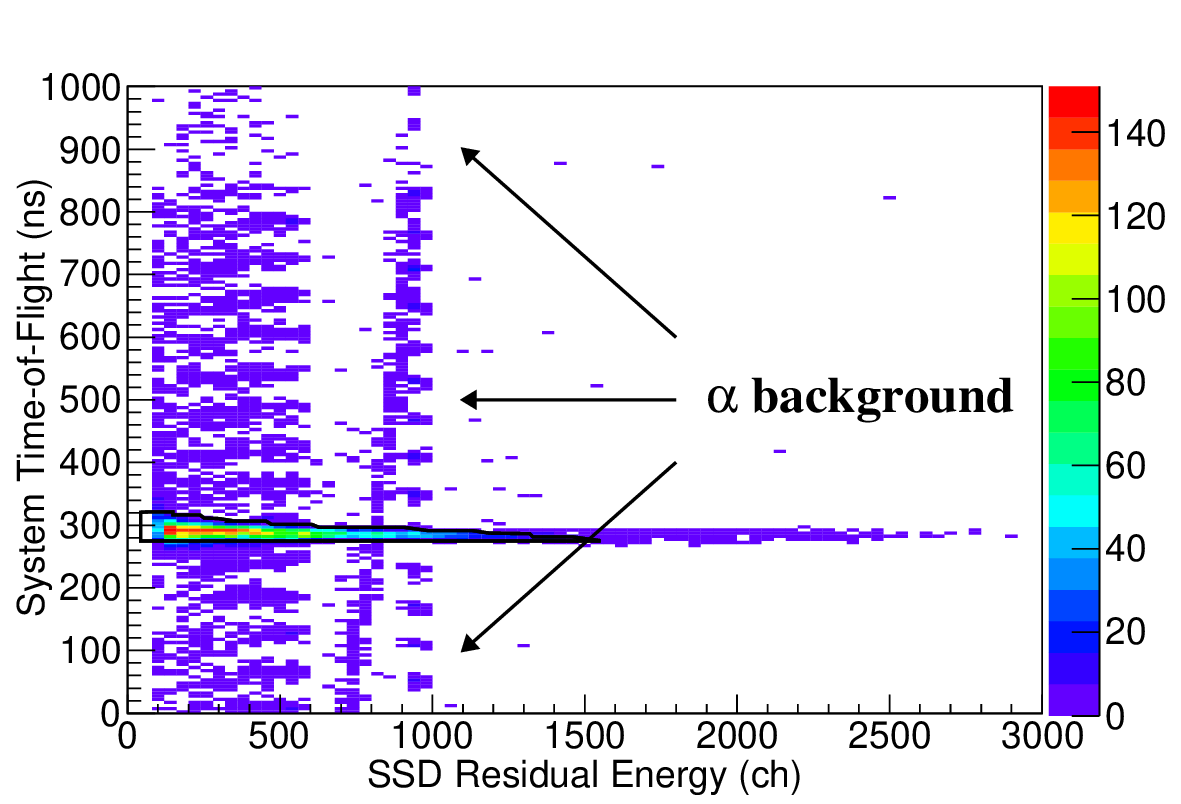} 
\caption
{
Residual light ion energy as measured by the SSD in channels on 
the abscissa against the system ToF in nanoseconds on the ordinate.  
Significant $\alpha$ background is seen around channel 1000 in the SSD energy.  
The locus of true elastic scattering events selected by the kinematic solution 
fall within the depicted gate; however, it can be observed that one locus of 
the beamlike $\alpha$ particles overlaps with the region of the true events.  
See the text.
}
\label{fig:background}
\end{figure}
The main sources of energy-dependent background could be $\alpha$ particles 
originating from the RIB production target satisfying the $B\rho$ selection as 
well as contributions from inelastic scattering. 
The bumps seen in the excitation function around 3.5~MeV in 
Fig.~\ref{rmatrix-good}(a) correspond to the region where $\alpha$ particles 
magnetically selected by CRIB are expected to appear.
These background ions are observed clearly in the spectrum of the SSD residual 
energy against the system time of flight (ToF) in Fig.~\ref{fig:background}.
The ToF is the time between PPACa and the SSD, following 
Ref.~\cite{2016PhRvC..93f5802H}.  
The figure shows all SSD events gated on incoming $^{30}$S ions, about $\approx 
80\%$ of which are discarded by the requirements of the kinematic solution.
The true elastic scattering events fall within a small locus on the histogram 
with a specific ToF, depicted by a narrow gate.
Conversely, the beamlike $\alpha$ particles span the entire ToF range with 
temporal spacing exactly corresponding to the inverse of the cyclotron 
radiofrequency signal, because these ions do not deposit enough energy to 
trigger the PPAC and merely arrive at the SSD in chance coincidence with a 
$^{30}$S ion at the PPAC. 
Ordinarily, the relation of the PPAC signal and cyclotron radiofrequency 
corresponds to a relative flight time of a beam ion within the cocktail beam 
(see Fig.~\ref{beam-pid2}), but this relation does not hold for such a chance 
coincidence.
Although most of such events are easily discarded, there is a small region of 
overlap where the beamlike $\alpha$ particles may contribute as much as 20\% of 
the data in this region.

There is another particle group to the left of the $\alpha$ background in 
Fig.~\ref{fig:background} which is quite unusual, as its residual energy is 
positively correlated with its system flight time.
The nature of this locus is not certain but is most likely associated with 
light ejecta resulting from heavy ions striking the Wien filter electrodes in 
chance coincidence with a $^{30}$S beam ion.
Considering the number of events in this locus per ns, we estimate it may 
contribute 5\% of the data in Fig.~\ref{rmatrix-good}(a) near 3.3~MeV.

As described below in Sec. \ref{sec:rmatrix}, introducing individual resonances 
in this region with widths equal to the theoretical limit made no discernible 
change to the calculated excitation function given our energy resolution.
Changing the fill-gas to perform background measurements with an active target 
necessitates adjusting the GEM high voltage settings (which would still not 
guarantee identical operating conditions) as well as replication of all tuning 
and calibration measurements, which requires a significantly larger investment 
of time compared to changing the target gas in an ordinary target.
Although we prepared to make such a measurement with an Ar+CO$_{2}$ gas mixture 
(as in our previous measurements, e.g., \cite{2013PhRvC..87c4303Y}), 
unfortunately we did not have enough time. 
Using the Wien filter to steer the beam, we determined that a vast majority of 
the beamlike $\alpha$ particles are confined to a narrow energy region.  

Further analysis of Fig.~\ref{fig:background} also sheds some light on the 
origin of the scaling factor of 2 applied to obtain the correct magnitude of 
the absolute value of the elastic scattering differential cross section.
As stated above, about 20\% of the events fall within a narrow locus and are 
consistent with bona fide elastic scattering events according to the kinematic 
solution algorithm.
More than three times this many events ($\sim60\%$ of the entire spectrum) fall 
within the depicted gate, but over 70\% of the events within this gate have no 
corresponding high-gain GEM data ($>40\%$ of the entire spectrum), thus 
$\vartheta_{\rm lab}$ is unknown and Eq.~\ref({eq:kinematics}) cannot be solved.
The solid angle of the right side of the downstream high-gain GEM is about 50\% 
that of the SSD, and the Rutherford cross section is known to go as 
$(\sin^{4}\frac{\theta}{2})^{-1}$, indicating it does not change appreciably 
over a small change ($\approx 10^{\circ}$) in $\vartheta_{\rm lab}$.
This implies that about $\frac{1}{2}$ of the events in the ToF-E locus of 
Fig.~\ref{fig:background} should be in coincidence with the high-gain GEM, but 
only about $\frac{1}{4}$ actually have high-gain GEM data.
These results are quite consistent with our scaling factor of 2.

As for possible contributions from inelastic scattering, the first excited 
state of $^{30}$S is relatively high at $E_{1{\rm x}}=2.21$~MeV and with a 
spin-parity of $2^{+}$.
The increased scattering threshold as well as the requirement for $\ell \geq 2$ 
from the angular momentum selection rules indicates that the widths, which 
decrease with increasing $\ell$, suggest a 
significantly lower cross section than elastic scattering.
For example, in other studies of $\alpha$ elastic scattering, this contribution 
was found to be less than 10\% \cite{2011PhRvC..83c4306Y, 2013PhRvC..87c4303Y}, 
where the first excited states are much lower in energy.
Moreover, as the resonances we analyzed were in the region of $4.0 \leq E_{\rm 
c.m.} \leq 5.6$~MeV, contributions from inelastic scattering would show up near 
$1.8\leq E_{\rm c.m.} \leq 3.4$~MeV in the elastic spectrum, where resonances 
were neither resolved nor analyzed in our data.
Therefore, it is reasonable to neglect any possible contribution from inelastic 
scattering in the present analysis; thus, we assume the total width can be 
expressed
as $\Gamma = \Gamma_{\alpha_{0}}+\Gamma_{\rm p}$.

\subsection{Experimental error}\label{sec:uncert}

\begin{figure}
 \includegraphics[scale=0.4, trim= 0 0 0 0]{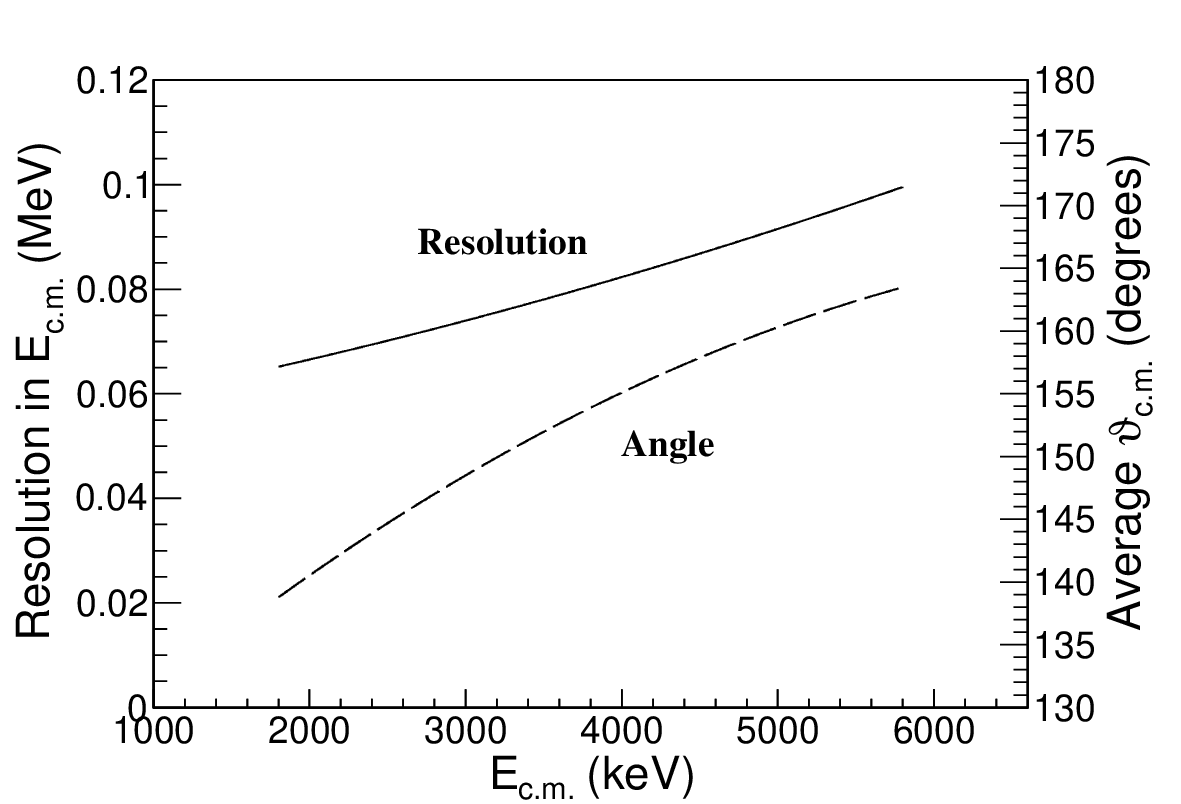} 
\caption
{
Uncertainty in determination of the center-of-mass energy $E_{\rm c.m.}$ in~MeV 
(solid line) and average scattering angle $\vartheta_{\rm c.m.}$ in degrees 
(dashed line) as functions of the center-of-mass energy in keV.  The range of 
angles measured in the laboratory frame is roughly $10^{\circ}\leq 
\vartheta_{\rm lab} \leq 20^{\circ}$ as $2\vartheta_{\rm lab} + 
\vartheta_{\rm c.m.}=180^{\circ}$.
}
\label{fig:resolution}
\end{figure}

A number of different factors can influence the determination of the 
center-of-mass energy $E_{\rm c.m.}$ for a given event: the spread in the beam 
energy from the momentum selection as well as straggling, the SSD resolution 
for measurement of the $\alpha$ particle residual energy, the straggling of the 
$\alpha$ particle, and the position determinations of both the recoiling 
$\alpha$ and beam ion.
However, since we use the geometric measurements to determine $\vartheta_{\rm 
lab}$ and the residual energy of the outgoing $\alpha$ particle to deduce 
$E_{\alpha}$, these have the most profound effect on the determination of 
$E_{\rm c.m.}$.
Based on Eq. (\ref{eq:kinematics}), the uncertainty in the center-of-mass energy 
$\Delta E_{\rm c.m.}$ can be expressed as
\begin{equation}
\frac{\Delta E_{\rm c.m.}}{E_{\rm c.m.}}=\sqrt{ \Bigg(\frac{\Delta 
E_{\alpha}}{E_{\alpha}}\Bigg)^{2} + 4\Bigg(\frac{\cos(\vartheta_{\rm 
lab})-\cos(\vartheta_{\rm lab}^{\prime}) }{\cos(\vartheta_{\rm 
lab})}\Bigg)^{2}},
\end{equation}
where $\Delta E_{\alpha}$ is the uncertainty in the measured $\alpha$-particle 
energy, $\vartheta_{\rm lab}$ is the average measured angle, and the 
uncertainty in the measured angle is $\Delta \vartheta_{\rm lab} = 
|\vartheta_{\rm lab}-\vartheta_{\rm lab}^{\prime}|$.
In the following illustrative calculations, $E_{\rm c.m.}$ was varied in 1~MeV 
increments over the range of 2--6~MeV.

Under the experimental conditions, the energy resolution of the SSD for 4.78-, 
5.48-, and 5.795-MeV $\alpha$ particles from a standard source was 103, 98, and 
87~keV, respectively.
For higher energy $\alpha$ particles, we assumed the resolution of 1.5\% as 
measured for the 5.795-MeV $\alpha$ particles, which should be an overestimate.
In an offline test, the SSD resolution for the 5.48-MeV line was as good as 
29~keV under vacuum which broadened to 70~keV when the chamber was filled with 
He+CO$_{2}$ gas; by folding an assumed 64~keV of broadening from energy 
straggling with the intrinsic SSD resolution, we were able to reproduce the 
measured width.
Considering the position of the $\alpha$ source was nearly 40 cm from the SSD 
in offline tests and $\alpha$ particles scattered at an initial laboratory 
energy of 5.5~MeV would be nearly twice as close to the SSD, 64~keV can be 
considered the maximum uncertainty for straggling, with higher energy $\alpha$ 
particles straggling much less as well as originating much closer to the 
detectors.
We finally adopted values for $\Delta E_{\alpha}$ by adding the above SSD 
resolution and the assumed straggling in quadrature, except for the highest 
energy $\alpha$ particles where we simply adopted an uncertainty of 1.5\% since 
summing the overestimated uncertainties from both resolution and straggling 
effects is unreasonable.

In order to estimate the uncertainty in $\vartheta_{\rm lab}$ arising from the 
experimental determination of the scattering position, we need to first 
estimate the average $\vartheta_{\rm lab}$ as a function of $E_{\rm c.m.}$.
We plotted both the laboratory scattering angle $\vartheta_{\rm lab}$ and the 
center-of-mass angle $\vartheta_{\rm c.m.}$ event by event in order to 
determine their average values as functions of the center-of-mass energy; the 
average $\vartheta_{\rm c.m.}$ is shown in Fig.~\ref{fig:resolution}.
While the precision of each PPAC to determine a beam particle's position is 
0.9~mm in both $X$ and $Y$, the position resolution becomes 4~mm in both 
dimensions when extrapolated to a typical scattering depth.
The resolution achieved for the $\alpha$ particle's position with the 
backgammon pads was 3~mm in $X$ and 0.5~mm in $Y$.
All these uncertainties were added together in quadrature to yield a final 
uncertainty of 6.4~mm in the determination of $\vartheta_{\rm lab}$.
A new angle $\vartheta_{\rm lab}^{\prime}$ was calculated by shifting the 
position of the $\alpha$ particle by the above 6.4~mm, assuming a standard 
scattering depth $Z$ representative of each of the five center-of-mass energies.
The resulting range of $\Delta \vartheta_{\rm lab}$ was found to be 
1.3--$2.1^{\circ}$, increasing with decreasing $E_{\rm c.m.}$.

Finally, we obtained an estimate for the uncertainty of the center-of-mass 
energy of about 60--100~keV as shown in Fig.~\ref{fig:resolution}; the 
intrinsic resolution of the SSD had the predominant effect, which was more 
pronounced at the higher energies.
Thus, it can be seen that the energy binning choice of 100~keV is consistent 
with our achieved resolution.

We confirmed with a simple calculation that the above geometric uncertainties 
dominate over the uncertainty in the beam energy.
Suppose we have two identical measurements, but we know that the incident 
energy differs between the two beam ions.
The result of the kinematic solution is that the optimized scattering depth 
will be larger for the higher energy beam ion and vice versa for the low energy 
beam ion, because it is the scattering depth combined with the incident beam 
energy together that finally determines $E_{\rm c.m.}$.
Assuming a nominal scattering energy of $E_{\rm c.m.}=4.0$~MeV, changing the 
transverse scattering position by the 6.4~mm uncertainty mentioned above is 
equivalent to $\Delta \vartheta_{\rm lab}=1.7^{\circ}$, changing the 
scattering depth $\Delta Z$ by 27~mm, or changing the beam energy by 5.6~MeV.
Thus, the uncertainties of these measurements dominate over the intrinsic 
spread in the beam energy of 2.0~MeV.

\subsection{$R$-Matrix analysis}\label{sec:rmatrix}
\begin{figure*}
    \subfigure{\includegraphics[angle=0, trim=0 0 0 
0,scale=.4]{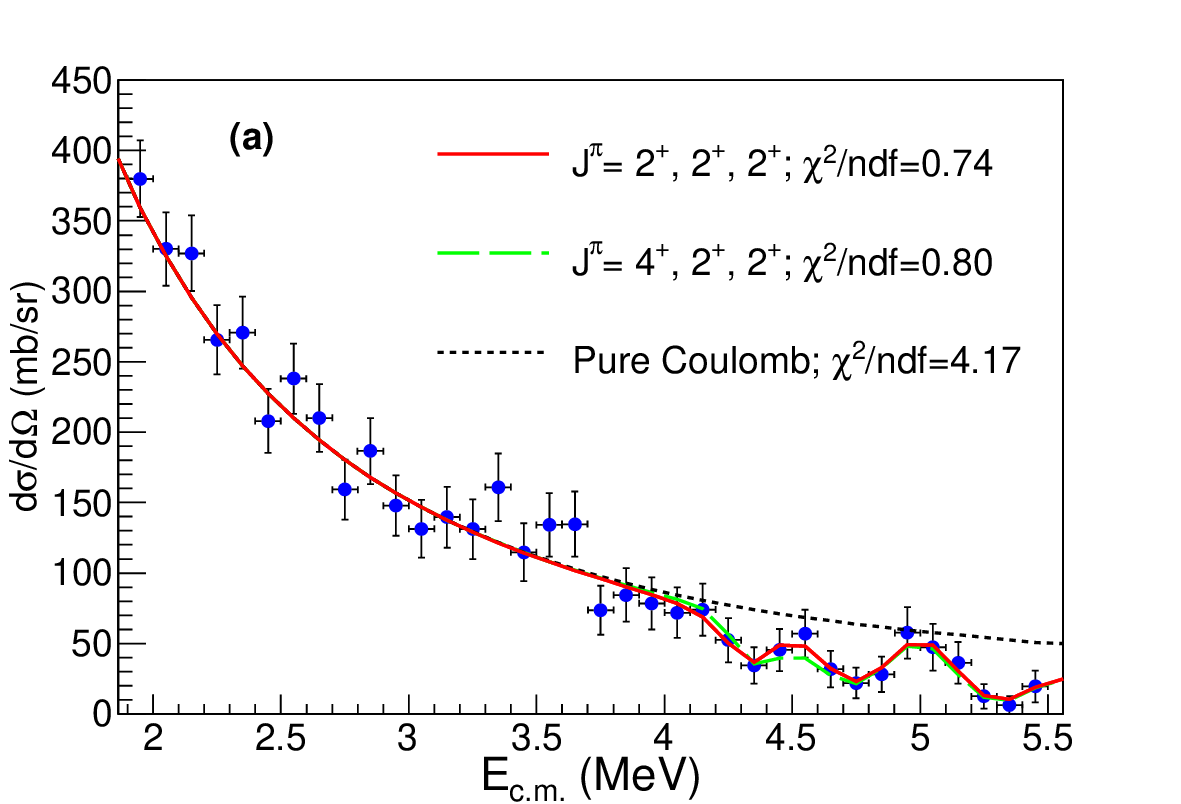}}
    \subfigure{\includegraphics[angle=0, trim=0 0 0 
0,scale=.4]{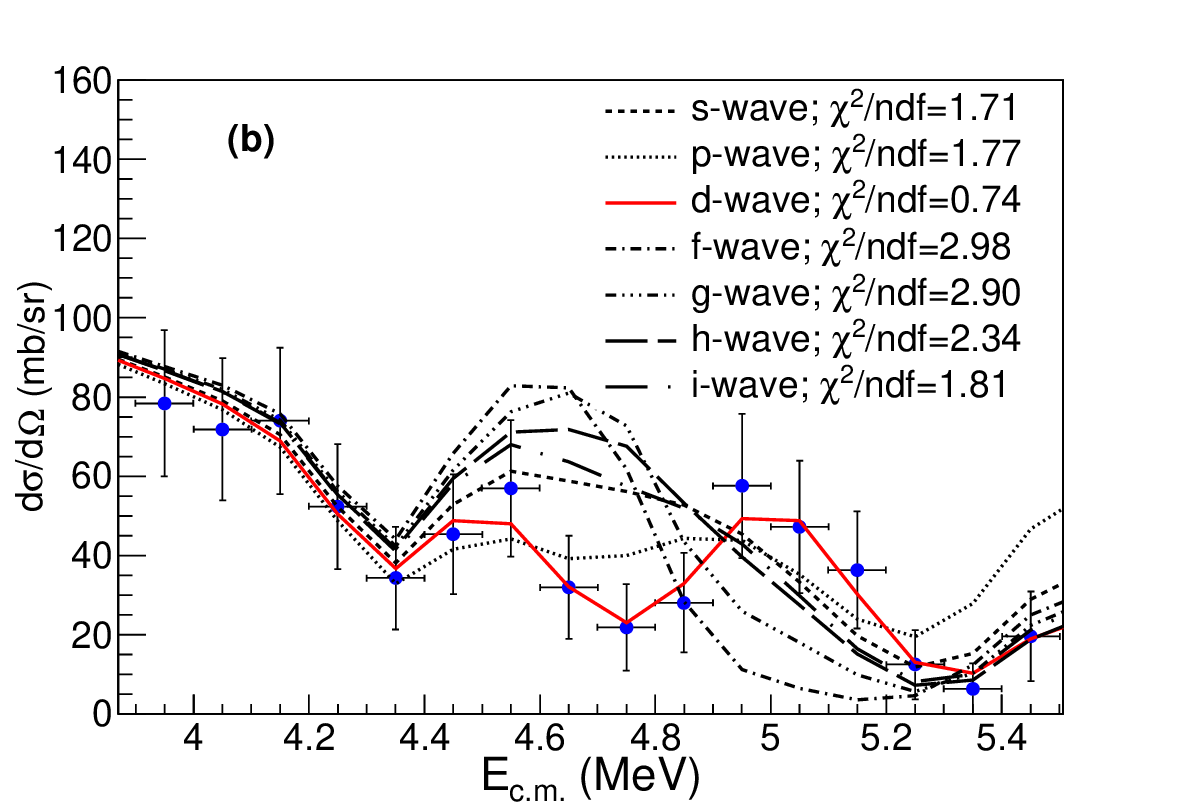}}
      \caption[Adopted $R$-Matrix fit]{
      $^{30}$S+$\alpha$ elastic scattering excitation function 
including fits.
      (a) The energy range displayed is the entire set of continuous data in 
the raw excitation function, except at the lower energy side where the plot is 
terminated at the point where all the $\alpha$ particles can no longer reach 
the detector from stopping in the fill gas.
      The bumps observed around 3.5~MeV correspond to a region of large 
$\alpha$ background, as depicted in Fig.~\ref{fig:background}.
      Three resonancelike structures are seen at $4.0$ MeV 
      $< E_{\rm c.m.} < 5.5$~MeV.
      The data are fit with a multichannel ($\alpha$ and p), multilevel 
$R$-Matrix formalism, and the results for a selected combination of 
$\ell_{\alpha}$ transfers are shown (though all combinations up to 
$\ell_{\alpha}\leq4$ were tested, and $\ell_{\alpha}=5,6$ never gave good 
fits).  The adopted parameters of these three newly discovered resonances are 
shown in Table \ref{tab:resparams}.
      (b) All physically allowed $\ell_{\alpha}$ values for the $E_{\rm 
r}=4.78$ MeV resonance, showing the unambiguous assignment of $\ell=2$.
      See the text.
      }
      \label{rmatrix-good}
      \end{figure*}

 \begin{table}
 \caption{\label{tab:ellp}
Coupling schemes for states in $^{34}$Ar for $J^{\pi}\leq4^{+}$ for the 
$^{33}$Cl+p channel.  The lowest $\ell_{\rm p}$ is assumed, and not all 
possible linear combinations are denoted.  See the text. 
}
 \begin{ruledtabular}
 \begin{tabular}{ccccc}
      $J^{\pi}$ & $\ell_{\rm p}$ & $s$ & $s_{1}\oplus s_{2}$ & $s\oplus\ell$  
\\ \hline
      $0^{+}$   & 2    &  2 & $\uparrow\uparrow$ & $\uparrow\downarrow$  \\  
      $1^{-}$   & 1    &  2 & $\uparrow\uparrow$ & $\uparrow\downarrow$ \\ 
      $2^{+}$   & 0    &  2 & $\uparrow\uparrow$ & --- \\ 
      $3^{-}$   & 1    &  2 & $\uparrow\uparrow$ & $\uparrow\uparrow$   \\ 
      $4^{+}$   & 2    &  2 & $\uparrow\uparrow$ & $\uparrow\uparrow$   \\ 
 \end{tabular}
 \end{ruledtabular}
 \end{table}

To extract the resonance parameters of interest, we performed a multilevel, 
multichannel $R$-Matrix calculation \cite{1958RvMP...30..257L} with the {\sc 
sammy8} code \cite{sammy8}.
Succinctly, the $R$-Matrix method calculates the interference between the 
regular and irregular Coulomb functions with physical resonances.
The resonances are parameterized by their energy $E_{\rm r}$ (the same as 
$E_{\rm c.m.}$ from elastic scattering as $Q=0$), channel $i$ partial widths 
$\Gamma_{i}$, and the angular momenta transfer $\ell_{i}$.
The resonance shape is determined by the entrance channel $\ell_{\alpha}$, the 
resonance height from the entrance channel $\Gamma_{\alpha}$, and the resonance 
width depends on total width $\Gamma$.
The total width is a sum of the proton and $\alpha$ partial widths, as both 
channels are open; the gamma partial widths $\Gamma_{\gamma}$ are negligibly 
small for these highly excited, particle-unbound states. 
For the case of $^{30}$S+$\alpha$ elastic scattering, the situation is 
simplified for the entrance channel, as both the nuclei have a ground-state 
spin-parity $J^{\pi}=0^{+}$, and so the quantum selection rules dictate a 
unique resonance $J^{\pi}$ for each $\ell_{\alpha}$ value---namely that 
$J=\ell_{\alpha}$ and the parity is always natural for populated states in 
$^{34}$Ar.

The calculated excitation function was broadened based on the experimental 
energy resolution and performed at an average angle of $\vartheta_{\rm 
c.m.}=150^{\circ}$ as evaluated above in Sec. \ref{sec:uncert}.
We quantified the quality of a fit by the reduced chi-square $\chi^{2}_{\nu}$, 
which is the chi-square $\chi^{2}$ divided by the number of degrees of freedom 
$\nu$.
Fitting the data with pure Coulomb scattering resulted in $\chi^{2}_{\nu}=4.17$ 
with 35 degrees of freedom, indicating the possibility for significant 
improvement could be expected by including the interference effect of 
resonances in an $R$-Matrix fit.
As there are no known levels in $^{34}$Ar with $E_{\rm ex}>11$~MeV, we had to 
carefully introduce new resonances until the experimental data were reasonably 
reproduced.
The maximum width of a resonance can be estimated with the Wigner limit 
\cite{RR88} as 
\begin{equation}\label{eqn:wigner}
W_{\Gamma_{i}} = \frac{2\hbar^{2}}{\mu_{i} R_{i}^{2}} P_{\ell_{i}},
\end{equation}
where $\mu$ is the channel reduced mass, $R$ is the channel radius, and 
$P_{\ell}$ is the channel penetrability, respectively, for channel $i$.
We calculate the penetrability as 
$P_{\ell}=\frac{\rho}{F_{\ell}^{2}+G_{\ell}^{2}}$, where 
$\rho=\frac{kR}{\hbar}$ includes the phase space factor $k$, and $F_{\ell}$ 
and $G_{\ell}$ are the regular and irregular Coulomb functions, respectively.
Such a physical constraint is particularly relevant when introducing new 
resonances to help limit the parameter space.
We adopted the channel radius given by $R_{i}=1.45(A^{1/3}_{1}+A^{1/3}_{2})$ 
fm, where $A_{1}$ and $A_{2}$ are the mass numbers of the two species in 
channel $i$; an identical parametrization was used in the studies of 
$^{21}$Na+$\alpha$ \cite{binhphd} and $^{26}$Si+p \cite{2012PhRvC..85d5802J}, 
which are two of the most similar experiments to the present work.
For consistency, the same $\alpha$-channel radius was also used in the 
$R$-Matrix calculation.
 
\begin{table*}
\caption{\label{tab:resparams}
Best fit level parameters of $^{34}$Ar determined by the present work.  All 
levels are newly proposed.  The table is arranged such that the corresponding 
physical property of each state in $^{34}$Ar precedes the corresponding 
$R$-Matrix fit parameter.   As we could not uniquely constrain the spin-parity 
of the 11.09-MeV level, two possible assignments are given, as well as the 
corresponding widths.  The 12.08-MeV level is shown in italic letters as there 
is a large systematic uncertainty associated with it. See the text.
}
 \begin{ruledtabular}
 \begin{tabular}{cccccccc}
        $E_{\rm ex}$ (MeV) &$E_{\rm r}$ (MeV) & $J^{\pi}$  &  $\ell_{\alpha}$ & 
 $\Gamma_{\alpha}$ (keV)  & $\theta^{2}_{\alpha}$ (\%) & $\Gamma_{\rm p}$ (keV) 
& $\xi$ (\%) \\ \hline
$11.092(85)$ & $4.353(85)$               & $(2^{+}, 4^{+})$ & 2, 4 & 
$20_{-17}^{+80}$, $0.5_{-0.4}^{+1.4}$   & $40_{-33}^{+180}$, $8_{-6}^{+10}$   & 
25$_{-20}^{+500}$, $0.3_{-0.3}^{+3.5}$ 
& 1, 0.1 \\ 
$11.518(89)$ & $4.779(89)$               & $2^{+}$        & 2      & 
$100_{-60}^{+120}$   & $90_{-55}^{+110}$   & 210$_{-170}^{+600}$ & 4  \\ 
\textit{12.079(95)} & \textit{5.340(95)} & \textit{(2\textsuperscript{+})}      
  & \textit{2}    & \textit{260}$_{-120}^{+400}$  & \textit{100}$_{-45}^{+150}$ 
& \textit{340}$_{-200}^{+550}$ & \textit{6}   \\ 
 \end{tabular}
 \end{ruledtabular}
 \end{table*}

At the outset, we began with a single channel ($\Gamma=\Gamma_{\alpha}$), 
single level manual analysis starting with the lowest-energy features and 
slowly moving to higher excitation energies in discrete steps of 100~keV.
The width was set to the Wigner limit ($\Gamma_{\alpha}=W_{\Gamma_{\alpha}}$) 
to determine which features could be resolved by assuming the existence of a 
physical resonance.
At this time we also checked possible values of the angular momentum transfer 
$\ell_{\alpha}$; although the experimental setup allowed for values of up to 
$\ell_{\alpha}=6$, $\ell_{\alpha}\geq 5$ never gave good fits, since resonances 
with these higher transfers are essentially not visible within the present 
resolution.
Only under this condition where $\Gamma_{\rm p}=0$ and $\Gamma_{\alpha}$ was at 
the Wigner limit was it possible to observe a change of any kind in the 
calculation for $E_{\rm c.m.}\leq 3.8$~MeV, and even so the calculated 
deviation from pure Coulomb scattering was of a smaller magnitude than the 
experimental uncertainty, particularly near 3.5~MeV.
The calculations were consistent with our interpretation that the fluctuations 
below 3.8~MeV are statistical or background induced.
The subsequent multilevel, multichannel analysis thus focuses on the region of 
3.9--5.6~MeV and assumes $\ell_{\alpha}\leq 4$; three resonancelike structures 
could be resolved near $E_{\rm c.m.}\approx 4.35$, 4.78, and 5.34~MeV.
Although resonances observed by transfer reactions always appear as peaks in 
the differential cross section, in the case of elastic scattering the 
interference pattern caused by a resonance can be observed as a diplike 
structure rather than as a peak, particularly below the Coulomb barrier.

For the proton channel, we assumed the lowest $\ell_{\rm p}$ allowed would have 
the predominant contribution.
The spin of the proton $s_{1}=\frac{1}{2}$ and the spin of the $^{33}$Cl ground 
state $s_{2}=\frac{3}{2}$, which can align ($\uparrow\uparrow$) or anti-align 
($\uparrow\downarrow$) to give the total spin $s=s_{1}\oplus s_{2}$, and the 
same is true for the resulting spin $s$ coupling with $\ell_{\rm p}$ to sum 
$J=\ell_{\rm p}\oplus s$.
An example of the lowest-$\ell_{\rm p}$ coupling schemes are shown in Table 
\ref{tab:ellp} for up to $4^{+}$ natural-parity states in $^{34}$Ar.

For convenience, we introduced the dimensionless reduced partial width 
$\theta_{i}^{2}=\Gamma_{i}/W_{\Gamma_{i}}$, in order to easily ensure that, 
regardless of $\ell$, $\theta_{i}^{2}\leq 1$.
Resonant elastic scattering is often analyzed by a single-channel formalism 
because the resonance shape and height are not affected by the other channels; 
thus at the outset we simplified our model by controlling the proton width via 
a universal ratio of the dimensionless reduced partial widths 
$\xi\equiv\theta_{\rm p}^{2}/\theta_{\alpha}^{2}$, which was found to be 3\% in 
a similar work \cite{binhphd}.
Although the value of $\Gamma_{\rm p}$ derived this way may have a large 
uncertainty as well as model dependence, it is physically unrealistic to 
perform a single-channel analysis so far above the proton threshold.

Starting with the first resonance near 4.35~MeV and truncating the excitation 
function towards higher energies, a computer code optimized $E_{\rm r}$, 
$\ell$, $\theta_{\alpha}^{2}$, and $\xi$, until all three resonances were 
introduced and the fit took into account the entire energy range of the 
experimental excitation function.
Once we had such a reasonable fit ($\xi\approx4$\%), we then allowed 
$\theta_{\rm p}^{2}$ to vary individually for each resonance and again covaried 
sets of resonance parameters to search for the best fit for the entire spectrum.
In summary, $E_{\rm r}$ was covaried over 200~keV in 1~keV steps, $\ell$ was 
covaried for values over the range of 0--4, $\theta_{\alpha}^{2}$ was 
covaried in 1\% steps up to $>99$\%, and $\theta_{\rm p}^{2}$ was varied in 
small increments up to 10\% (past where $\xi$ showed poor behavior) in our 
search for the best fit, shown in Fig.~\ref{rmatrix-good}(a), where the 
horizontal errors are from the 100-keV binning and the vertical errors are 
purely statistical; the absolute magnitude of the statistical error was scaled 
by the same factor of 2.0 as the data.
All possible $\ell$ values for the 4.78-MeV state are shown in 
Fig.~\ref{rmatrix-good}(b) for illustrative purposes.

A number of systematic uncertainties that might affect the differential 
cross section were carefully considered.
These include contributions to each of the parameters in 
Eq.~\ref({eq:dsigma-domega}), namely the number density of helium atoms $n$, the 
stopping power of the beam $S(E_{\rm b})$, the number of injected beam ions 
$I_{\rm b}$, and the changing solid angle $\Delta\Omega$.
The number density of $^{4}$He atoms in the target employed cannot be changed 
by any physical argument, since the pressure gauge in the gas flow controller 
described in Sec.~II was consistent with two other pressure gauges.
The gas density calculated from the nominal laboratory conditions with the 
ideal gas law is consistent with the density utilized in all energy loss 
calculations.
Any error in the density of helium would apply equally to CO$_{2}$ which 
induced significant energy loss for the heavy ions.
Such energy loss calculations are generally known to be accurate to the order 
of 10\% or better, and moreover we experimentally verified the stopping 
position and Bragg curve of $^{30}$S in the target gas.
Although we initially assumed the energy distribution of the $^{30}$S ions 
should be gaussian, we found the centroid was skewed to the low energy side by 
1\%; calculating the $^{30}$S ion incident energy event-by-event based on the 
$rf$ data rather than assuming a gaussian distribution did not result in any 
noticeable difference in the elastic scattering excitation function.
The number of incident $^{30}$S ions is determined in part by the PPAC scalers, 
the magnitude of which can be confirmed as we recorded a downscaled spectrum of 
the cocktail beam; although the latter is less accurate, the two methods agreed 
within 6\%.
We checked the method of calculating the solid angle, as well as the absolute 
efficiency of the silicon detectors, considering the known intensity of the 
standard $\alpha$ source used in off-line calibration runs; within the errors of 
these calculations, we found the efficiency $\eta>99\%$ for the relevant 
silicon detectors before and after the experimental run, indicating the 
detectors were not damaged during the experiment.
As the scaling factor of $2.0\pm0.1$ is the main source of systematic error, we 
take its associated uncertainty of 5\% as the systematic uncertainty in the 
present work.
Because the statistical error is $\geq25$\% over the resonant-dominated region, 
the systematic error makes a negligible contribution to the final error 
evaluation.

The resonance parameters deduced from the $R$-Matrix analysis are shown in 
Table \ref{tab:resparams}.
The uncertainties in the adopted level parameters were calculated in the 
following ways.
For the excitation energy $E_{\rm ex}$, we used the experimental energy 
resolution as discussed in Sec. \ref{sec:uncert} and shown in 
Fig.~\ref{fig:resolution}.
The error of the remaining level parameters was evaluated considering the range 
where an individual parameter is allowed to vary within one standard deviation 
of the best fit $\chi^{2}_{\nu}$; the same method was used over the range of 
$E_{\rm c.m.}$ 1.9--4.1~MeV with pure Rutherford scattering to determine the 
scaling factor and its error of $2.0\pm0.1$. 
The recommended spin parity $J^{\pi}$ is given, and any other spin-parity which 
is possible is listed, as are the associated widths in their respective columns 
separated by commas. 
The error in $\Gamma_{\rm p}$ is seen to be generally larger than in 
$\Gamma_{\alpha}$, because the $\alpha$ elastic scattering resonant structure 
is much less sensitive to the proton channel compared to the $\alpha$ channel.

The resonance parameters obtained in the present study appear to be reasonable 
except for the widths for the 12.08-MeV state.
In particular, the 12.08-MeV state's structure cannot be a pure $\alpha$ 
cluster which also has a non-negligible proton decay branch.
Our favored interpretation is that there are one or more unresolved resonances 
with substantial $\alpha$-cluster configuration in this region.
Moreover, the behavior of the resonance tail is unconstrained by the data, and 
any interference effects from unknown physical resonances outside the energy 
range cannot be accounted for.
Thus, there are large systematic uncertainties associated with the resonance 
parameters extracted from an $R$-Matrix fit for states near the boundary of the 
experimental energy range.
However, it cannot be doubted that the data indicate one or more very strong 
$\alpha$-cluster resonance(s) in this region of excitation energy, which is a 
point we emphasize in our discussion of these results below.

\section{Discussion}

\begin{table*}
\caption{\label{tab:resdata}
Resonance parameters of $^{34}$Ar adopted in the calculation of the 
$^{30}$S($\alpha$,\,p) stellar reaction rate calculation.  
Resonances with $E_{\rm r}<0.7$~MeV are not tabulated as they fall below the 
x-ray burst Gamow window and the Wigner limit for the $\alpha$ channel rapidly 
vanishes.  
Parameters shown in boldface are based on experimental data.
Level energies below $E_{\rm ex}<11$~MeV are taken from 
Ref.~\cite{2009AIPC.1090..288O}, where we assumed $J^{\pi}=0^{+}$ and 
$\Gamma_{\alpha}=\frac{1}{2}W_{\Gamma{\alpha}}=\gamma$; 
the relative dependence 
of $J$ and $\gamma$ for small $J$ in our framework is exemplified in 
Fig.~\ref{fig:kahl_limit}.
The higher energy resonances are from the present work and are separated from 
the others by a horizontal rule.
We stress that the tabulated properties which are not taken from experimental 
data may not be correct for individual resonances, but rather that the sum of 
these contributions to the stellar reaction rate can be considered an upper 
limit under an extreme assumption, which is interesting to investigate.
See the text.
}
 \begin{ruledtabular}
\begin{tabular}{cccccccc}
      $E_{\rm ex}$ (MeV) & $E_{\rm r}$ (MeV) & $J^{\pi}$ & $\omega$  & 
$\Gamma_{\alpha}$ (keV) & $W_{\Gamma_{\alpha}}$ (keV) & $W_{\Gamma_{\rm p}}$ 
(MeV) & $\omega\gamma$ (keV) \\  
      \hline 
      \textbf{7.47}   & \textbf{0.73}  & ($0^{+}$) &1& 
$2\times10^{-15}$&$4\times10^{-15}$   & $2\times10^{-1}$  &  $2\times10^{-15}$  
   \\ 
      \textbf{7.88}   & \textbf{1.14}  & ($0^{+}$) &1& $2\times10^{-9}$ 
&$3\times10^{-9}$    & $4\times10^{-1}$  &  $2\times10^{-9}$ \\ 
      \textbf{7.96}   & \textbf{1.22}  & ($0^{+}$) &1& $1\times10^{-8}$ 
&$2\times10^{-8}$    & $4\times10^{-1}$  &  $1\times10^{-8}$ \\ 
      \textbf{8.15}   & \textbf{1.41}  & ($0^{+}$) &1& $4\times10^{-7}$ 
&$8\times10^{-7}$    & $6\times10^{-1}$  &  $4\times10^{-7}$ \\ 
      \textbf{8.30}   & \textbf{1.56}  & ($0^{+}$) &1& $4\times10^{-6}$ 
&$8\times10^{-6}$    & $6\times10^{-1}$  &  $4\times10^{-6}$ \\ 
      \textbf{8.55}   & \textbf{1.81}  & ($0^{+}$) &1& $1\times10^{-4}$ 
&$2\times10^{-4}$    & $8\times10^{-1}$  &  $1\times10^{-4}$ \\ 
      \textbf{8.74}   & \textbf{2.0 }  & ($0^{+}$) &1& $7\times10^{-4}$ 
&$1\times10^{-3}$    & $9\times10^{-1}$  &  $7\times10^{-4}$ \\ 
      \textbf{8.89}   & \textbf{2.15}  & ($0^{+}$) &1& $3\times10^{-3}$ 
&$5\times10^{-3}$    & 1  &  $3\times10^{-3}$  \\ 
      \textbf{8.99}   & \textbf{2.25}  & ($0^{+}$) &1& $7\times10^{-3}$ &0.01   
             & 1  &  $7\times10^{-3}$ \\ 
      \textbf{9.42}   & \textbf{2.68}  & ($0^{+}$) &1& 0.1              &2      
             & 1  &  0.1  \\ 
      \textbf{9.75}   & \textbf{3.01}  & ($0^{+}$) &1& 0.7              &1      
             & 2  &  0.7  \\ 
      \textbf{10.32}  & \textbf{3.58}  & ($0^{+}$) &1& 7                &10     
             & 2  &  7   \\ 
      \textbf{10.81}  & \textbf{4.07}  & ($0^{+}$) &1& 30               &50     
             & 3  &  30    \\ \hline 

\textbf{11.09} & \textbf{4.35} & \textbf{2\textsuperscript{+}} &\textbf{5} & 
\textbf{20 } & 50  & 5 & 100     \\ 
\textbf{11.52} & \textbf{4.78} & \textbf{2\textsuperscript{+}} &\textbf{5} & 
\textbf{100} & 110 & 5 & 500     \\ 
\textbf{12.08} & \textbf{5.34} & \textbf{2\textsuperscript{+}} &\textbf{5} & 
\textbf{260} & 260 & 6 & 1300    \\ 
 \end{tabular}
 \end{ruledtabular}
 \end{table*}

We observed the signature interference patterns of several resonances in 
$^{34}$Ar with large $\alpha$ partial widths $\Gamma_{\alpha}$ via $\alpha$ 
elastic scattering on $^{30}$S.
The cluster threshold rule predicts the existence of these states, which have a 
large overlap of the cluster configuration to the nuclear wavefunction nearby 
the respective cluster's separation energy 
\cite{1994ZPhyA.349..237K,2010NuPhA.834..647K}.
Such $\alpha$-cluster resonances have typically dominated the stellar rate of 
exothermic ($\alpha$,\,n) and ($\alpha$,\,p) reactions on $T_{z}=\pm1$ nuclei, 
respectively, when they fall within the astrophysical Gamow burning window 
\cite{2005PrPNP..54..535A}. 
$\alpha$ resonant elastic scattering has long been known as a powerful tool to 
selectively observe states with large $\Gamma_{\alpha}$.
The effect is especially pronounced in inverse kinematics, where measurements 
at large backward angles are possible and the nonresonant cross sections are 
minimized; under these conditions, one expects to observe states with 
$\Gamma_{\alpha}$ comparable to the experimental energy resolution 
\cite{1990SvJNP..52..408}.
According to calculations of the Wigner limit [see Eq. (\ref{eqn:wigner})], the 
maximum theoretical width shrinks rapidly as the energy is reduced towards the 
threshold.

Our observation that all three resonances are consistent with a $2^{+}$ 
assignment may make one wonder if there is a reason for such behavior.
These may be regarded as a triplet if the tentative assignments of $2^{+}$ are 
correct for both the 11.09- and 12.08-MeV states.
Alternatively, our 11.51-MeV state could be regarded as a $2^{+}$ doublet 
paired with either these other two states. 
We note that all three are observed as diplike structures, so it may not be 
surprising that features in the differential cross section with similar 
interference patterns can result from physical resonances with the same 
$J^{\pi}$.
A system of $\alpha$-cluster doublets was observed in the $T_{z}=+1$ 
nucleus $^{22}$Ne \cite{2001PhRvC..64e1302R,2004NuPhA.738..447D} with $J^{\pi}$ 
correlated with increasing energy, albeit for states of negative rather than 
positive parity.
Unfortunately, comparison with model predictions is still a challenge for the 
$^{30}$S mass region.

\subsection{Reaction rate}
\begin{figure}
 \includegraphics[scale=0.45, trim= 0 0 0 0]{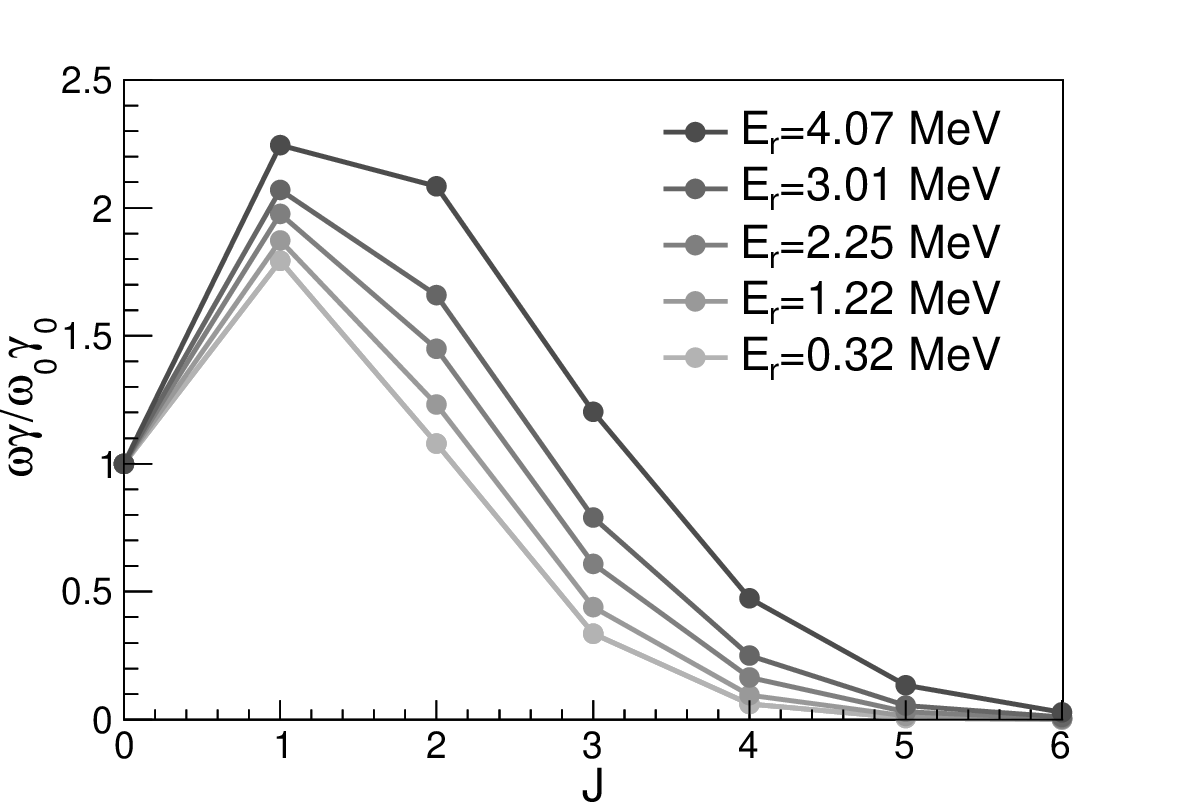}
\caption
{
Relative dependence of the resonance strength $\omega\gamma$ from individual 
states based on their spin $J$ under the assumption that 
$\gamma\approx\Gamma_{\alpha}\propto W_{\Gamma_{\alpha}}$.  The resonance 
strengths are normalized to the case of $J=0$, denoted as 
$\omega_{0}\gamma_{0}$.  The $E_{\rm r}=2.25$~MeV state is the closest to the 
center of the 1.3~GK Gamow window.
}
\label{fig:kahl_limit}
\end{figure}
The peak temperature of x-ray bursts is expected to be in the range of 1.3--2 
GK corresponding to the Gamow burning windows of $1.7 \lesssim E_{\rm c.m.} 
\lesssim 3.8$~MeV.
To make a meaningful evaluation of the stellar reaction rate in XRBs, we 
therefore need to consider not only the resonances discovered in the present 
work, but also $^{34}$Ar states at lower $E_{\rm ex}$.
In fact, before the present work there has never been an evaluation of the 
$^{30}$S($\alpha$,\,p) cross section based on experimental level structure of 
$^{34}$Ar owing to the paucity of such data and the experimental challenges of 
studies in this region of the periodic table.

The $^{36}$Ar(p,\,t)$^{34}$Ar measurement performed at the Research Center for 
Nuclear Physics (RCNP), Osaka University observed resonances above the $\alpha$ 
threshold at a relatively smooth interval---about four resonances per 
MeV---over four MeV in excitation energy \cite{2009AIPC.1090..288O}.
Considering the significant difference in energy resolution, our observation of 
three (or four, depending on the interpretation of our 12.08~MeV state) 
resonances per MeV over the range of our experimental energy and as the 
resolution allows, there is a basic consistency between the two studies, which 
both only populate natural-parity states.\footnote{While a single step (p,\,t) 
transfer reaction can only populate natural-parity states, a multistep process 
allows for the population of unnatural-parity states (see, e.g., 
Ref.~\cite{2013PhRvC..87f5801S}).  However, the cross section to populate 
unnatural-parity states by a multistep process is significantly smaller than 
the cross section for a single step process as a general rule.}

However, the results from the RCNP spectroscopic study only provides us with 
preliminary resonance energies, and some assumptions are required before we may 
apply them.
Firstly, we na\"ively assumed that each state has $J^{\pi}=0^{+}$.
As for the partial widths, based on the present results and the similar level 
density between the two studies, we set 
$\Gamma_{\alpha}=\frac{1}{2}W_{\Gamma_{\alpha}}$.
Although our $\theta_{\alpha}^{2}$ are generally larger than 0.5 according to 
Table \ref{tab:resparams}, setting $\theta_{\alpha}^{2}\approx1$ for such a 
large series of resonances would be unusual considering $\Gamma_{\rm p}\neq0$ 
and thus $\Gamma>\Gamma_{\alpha}$ and hence $\theta_{\alpha}^{2}<1$.
We believe a factor of 0.5 is still rather extreme but more reasonable.

It should be noted a thorough analysis of the RCNP experimental data would only 
improve the situation with regard to the precision of the excitation energies 
(or the removal of any states which are background induced) and not the 
spin-parities nor the partial widths.
The limited angular distribution available from their spectrometer does not 
cover a full phase for reliable comparison with a DWBA calculation 
\cite{2009PhRvC..80e5804M,2011PhRvC..84b5801M}. 
\begin{figure}
 \includegraphics[scale=0.45, trim= 0 0 0 0]{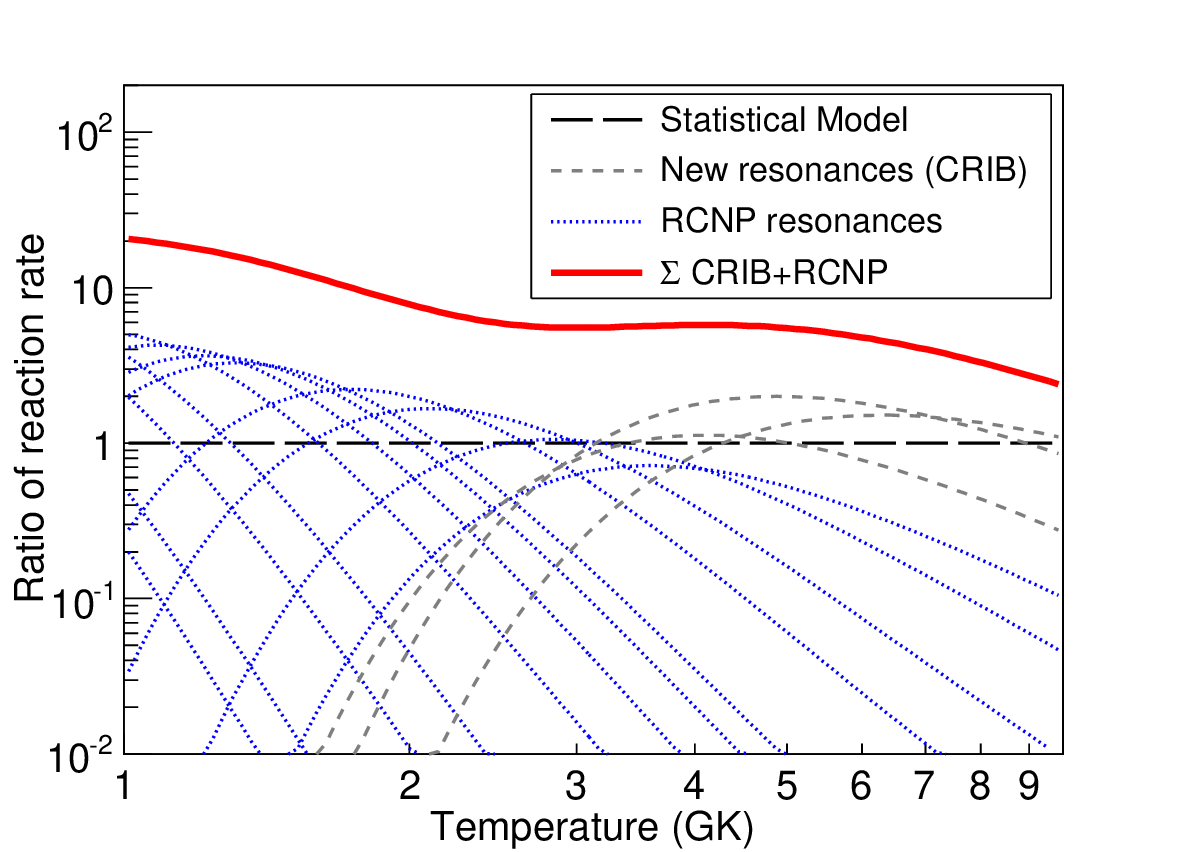}
\caption
{
Calculations of the $^{30}$S($\alpha$,\,p) stellar reaction rate 
from 1--10 GK. 
The statistical model (SM) rate from {\sc non-smoker} \cite{nonsmoker.web} is 
shown as the long-dashed black line, to which all the rates are normalized.
The dashed grey lines represent our new higher-energy resonances observed at 
CRIB with our best fit quantum properties.
The dotted (blue) lines represent the 13 resonances from the 
$^{36}$Ar(p,\,t) RCNP experiment \cite{2009AIPC.1090..288O}, where we made a 
couple of assumptions about their quantum properties.
The sum of these individual resonant contributions is shown as the thick solid 
(red) line.
The adopted individual resonance properties are listed in Table 
\ref{tab:resdata}.
}
\label{fig:rate}
\end{figure}

We calculated the resonant reaction rate per particle pair $\langle\sigma v 
\rangle$ using the standard formulation \cite{RR88} which depends only on the 
resonance energies $E_{\rm r}$, spins $J$, and the channel partial widths 
$\Gamma_{i}$.
The spin comes into play in the spin statistical factor $\omega$ as:
\begin{equation}
\omega \equiv \frac{2J_{\rm r}+1}{(2J_{\rm A}+1)(2J_{\rm a}+1)}~,
\end{equation}
where $J_{\rm A,a}$, the spins of the two nuclei in the entrance channel, are 
both zero in the case of $^{30}$S+$\alpha$.
The reduced width $\gamma$ is defined as:
\begin{equation}
\gamma \equiv \frac{\Gamma_{\rm a} \Gamma_{\rm b}}{\Gamma_{\rm tot}},
\end{equation}
for the entrance and exit channel partial widths $\Gamma_{\rm a,b}$, 
respectively.
Their product $\omega\gamma$ is called the resonance strength as it is 
proportional to the integral of the resonance cross section. 
We also use the standard simplification that $\gamma = (\Gamma_{\alpha} 
\Gamma_{p})/(\Gamma_{\alpha}+\Gamma_{p}) \approx \Gamma_{\alpha}$ when 
$\Gamma_{\alpha} \ll \Gamma_{p}$, which is a realistic assumption considering 
the vastly different Wigner limits of the two channels.\footnote{According to 
our data from Table \ref{tab:resparams}, the best fit $\Gamma_{\rm p}$ and 
$\Gamma_{\alpha}$ are of comparable size.  However, we note that the 
upper-limit errors for $\Gamma_{\rm p}$ are extremely large, which is not 
inconsistent with the assumption that $\gamma \approx \Gamma_{\alpha}$.  Even 
in the case that $\Gamma_{\alpha}=\Gamma_{\rm p}$, $\gamma 
=\frac{1}{2}\Gamma_{\alpha}$, giving a similar factor of 2 error to 
$\omega\gamma$ as the assumption of $J=0$ as shown presently.}
The resulting $\omega\gamma$ for each resonance in the $^{36}$Ar(p,\,t) study 
based on our assumptions that $J=0$ and 
$\gamma=\Gamma_{\alpha}=\frac{1}{2}W_{\Gamma_{\alpha}}$ varies by only a factor 
of around two for $J\leq 3$ (although it quickly drops off for $J\geq5$) as 
shown in Fig~\ref{fig:kahl_limit}, vindicating our arbitrary treatment of the 
spin 
in our framework; $\omega\gamma$ is independent of low $J$ to first order 
because we parametrize the width based on the Wigner limit, which decreases 
with increasing $J$, whereas for the $^{30}$S+$\alpha$ entrance channel 
$\omega=2J_{\rm r}+1$.
While the properties of an individual resonance calculated in this manner will 
be unreliable, the sum of these contributions can be considered an upper limit 
under extreme assumptions on the nuclear structure of $^{34}$Ar.

Our goal here is to provide an evaluation of the $^{30}$S($\alpha$,\,p) 
reaction rate in x-ray bursts by assuming broad widths for all the known
states in $^{34}$Ar near the astrophysically interesting region.
According to Fig.~\ref{fig:kahl_limit}, the choice of $J^{\pi}=0^{+}$ yields a 
median value of 
the resonance strength for low $J$ in our model and thus represents the 
physically 
realistic case where there are a multitude of different spins among the levels 
in $^{34}$Ar over the alpha threshold. 
It can be seen in the case of $J=3$ or higher that the reaction rate could be 
half or less than our suggested upper limit.
To demonstrate, we considered the case when $\gamma = \Gamma_{\alpha} = 
2W_{\Gamma_{\alpha}}$ and $J^{\pi}=4^{+}$; between $T = 1 and 3$~GK
the reaction rate is a factor of 2 {\it lower} than our evaluation in spite 
of the fact 
that the reduced width was twice rather than half the Wigner limit.
This feature of the Wigner limit is satisfyingly consistent with the fact that 
resonant thermonuclear reaction rates tend to be dominated by lower partial 
wave contributions because the angular momentum barrier is smaller.
Importantly, the rate only exceeds our evaluation by 0.2\% at $T=1.5$ GK if we 
allow the maximum $\Gamma_{\alpha}$ from our experimental uncertainties in 
Table II, which we consider to be a trivial difference. 

The $^{34}$Ar resonance parameters adopted for our $^{30}$S($\alpha$,\,p) 
stellar reaction rate calculation are listed in Table \ref{tab:resdata} along 
with calculations of the proton and $\alpha$ Wigner limits.
We used the graph digitizing system {\sc gsys} from the Hokkaido University 
Nuclear Reaction Data Centre, which was developed specifically for extracting 
numerical nuclear data from published spectra, to obtain the resonance energies 
from the $^{36}$Ar(p,\,t) study \cite{2009AIPC.1090..288O}.
For the known resonances, we found deviations from the compiled level scheme 
\cite{1990NuPhA.521....1E} of at most around 30 keV, which is the same as the 
experimental error quoted by O'Brien {\it et al.}
The resulting stellar reaction rate is shown in Fig.~\ref{fig:rate} in 
comparison to a statistical model (SM) rate \cite{nonsmoker.web}. 
As this is the first $^{30}$S($\alpha$,\,p) reaction rate based on the experimental 
level structure of $^{34}$Ar, there are not many other studies to compare ours 
against.
The two most relevant studies are the $^{33}$Cl(p,\,$\alpha$) measurement 
\cite{2011PhRvC..84d5802D} and a recent survey of $\alpha$-induced cross 
sections for masses $A\approx20$--50 \cite{2015EPJA...51...56M}.

In order to compare the present results with the time-reversal study by Deibel 
{\em et al.} \cite{2011PhRvC..84d5802D}, one should keep in mind that the 
$^{30}$S($\alpha$,\,p) $Q$ value is 2.080~MeV, and thus their energy range in 
$^{30}$S+$\alpha$ is $4.09 \leq E_{\rm c.m.}\leq 5.35$~MeV (as shown in their 
Table I), quite similar to the range of resonances observed in the present 
work. 
The previous study includes only the ($\alpha$,\,p$_{0}$) ground-state 
component, 
whereas the present $\Gamma_{\rm p}$ does not exclude the summed contribution
from all states where the assumption $\Gamma_{\alpha} \ll \Gamma_{\rm p_{i}}$ 
remains valid for at least one p$_{i}$;
the SM rate implicitly includes transitions to allowed states.
The present work shows an enhancement of around a factor of 5 over the SM 
rate as an upper limit, and the work of Deibel {\em et al.} shows a 
$^{30}$S($\alpha_{0}$,\,p$_{0}$) cross section which is more comparable to the 
SM rate and is considered as a lower limit to the total ($\alpha$,\,p) cross 
section. 
Even if one only includes the resonances we observed at CRIB, it can be seen 
that near 3 GK our three resonances alone are quite similar to the SM rate.

The reduction scheme presented by Mohr generally shows a global behavior of the 
cross sections for ($\alpha$,\,p) and ($\alpha$,\,n) reactions over a large 
energy range for medium mass nuclei \cite{2015EPJA...51...56M}.
Specifically, most of the experimental data can be reproduced by a SM 
calculation.
However, deviations higher than the expected cross sections were found in some 
of the measurements with $^{23}$Na and $^{33}$S, whereas the species $^{36}$Ar 
and $^{40}$Ar were seen to be much lower (at least for the available data).
For the case of $^{23}$Na, the work of Mohr motivated the community to
re-investigate the $^{23}$Na($\alpha$,\,p) cross-section, which was finally
found to be consistent with the SM calculation within the experimental 
uncertainties \cite{2014PhRvL.112o2701A, 2015PhRvL.115q9901A, 
2015PhRvL.115e2702T, 2015PhRvL.115e2701H}.
In the outstanding cases, these discrepancies certainly warrant further 
investigation to 
determine if they are real or artificial (see also the discussion in the 
recent work by Anderson {\em et al.} \cite{2017PhRvC..96a5803A}).
If the effects are real, $^{30}$S is seen to fall within the mass range where 
there is a cross section enhancement over the SM, which supports the findings 
of the present study.

\subsection{Astrophysical implications}
\begin{figure}
 \includegraphics[scale=0.55, trim= 0 0 0 0]{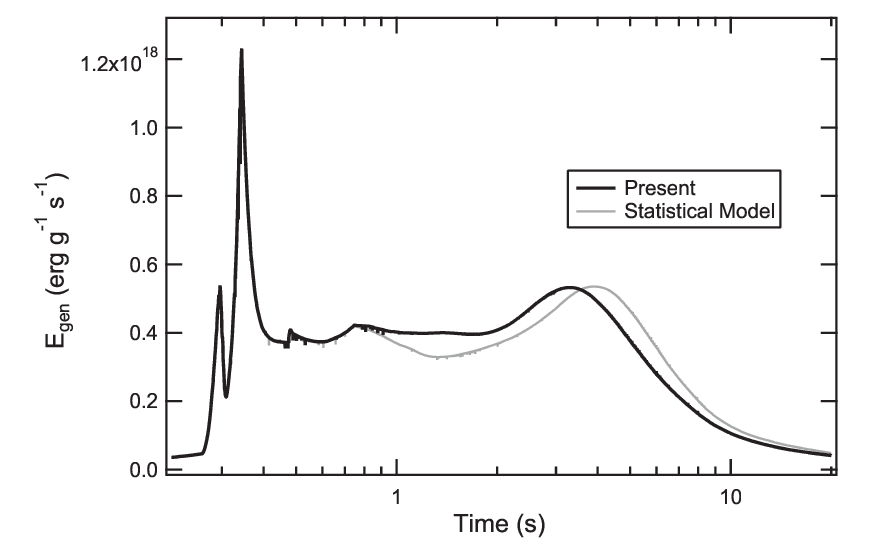}
\caption
{
Nuclear energy generation rates during one-zone XRB calculations using the {\tt 
K04} thermodynamic history \cite{2008ApJS..178..110P}. Results using the 
present rate (black line) and a statistical model rate \cite{nonsmoker.web} 
(grey line) are indicated.
}
\label{fig:egen}
\end{figure}
The impact of our new upper limit for the $^{30}$S($\alpha$,\,p)$^{33}$Cl rate 
was examined within the framework of one-zone XRB postprocessing calculations 
using the {\tt K04} ($T_{\rm peak}=1.4$~GK) model \cite{2008ApJS..178..110P}.  
As shown in Fig.~\ref{fig:egen}, striking differences in the profiles of 
nuclear energy generation rates between approximately 1 and 10 s are seen when 
comparing XRB calculations using the present upper limit and a statistical 
model rate calculation \cite{nonsmoker.web} (see Fig.~\ref{fig:rate}).  Indeed, 
$E_{\rm gen}$ differs by as much as 25\% between calculations using these two 
rates.  Nucleosynthesis predictions are also affected by the particular 
$^{30}$S($\alpha$,\,p) rate adopted in the calculations.  Comparing results 
using the present upper limit and the statistical model rate, abundance 
differences of up to 30\% are observed for species with mass fractions $> 
10^{-6}$ (summed over mass number), for $A$ over the rather broad range of 
approximately 20--$80$.  Further tests using full hydrodynamic XRB models are 
needed to explore in detail the possible dramatic impact of the 
$^{30}$S($\alpha$,\,p)$^{33}$Cl rate on predictions of XRB observables.

\section{Summary}
We observed several resonances with large $\alpha$ widths in the energy range 
$E_{\rm ex} =11.1$--12.1~MeV for the first time in $^{34}$Ar via the $\alpha$ 
resonant elastic scattering of $^{30}$S and determined their properties of 
spin, parity and widths.
Using our new data, we were able to make the first-ever calculation of the 
astrophysical $^{30}$S($\alpha$,\,p) cross section based on the experimental 
level structure of $^{34}$Ar.
Although these resonances do not seem to have a large effect for the 
astrophysically interesting energies important for XRBs, we could set a 
reasonable upper limit on the stellar reaction rate of about one order of 
magnitude greater than the Hauser-Feshbach statistical model.
The resonances we observed correspond very well to the energy range covered in 
the time-reversal study.
These two studies complement each other nicely, as our work provides an upper 
limit to the cross section, which we determined to be somewhat above the 
existing $^{33}$Cl(p$_{0}$,\,$\alpha_{0}$) lower limit to the total 
$^{30}$S($\alpha$,\,p) cross section.
Although the present knowledge of the level structure of $^{34}$Ar, as well as 
the $^{30}$S($\alpha$,\,p) cross section, at the most astrophysically 
interesting temperatures remains elusive, our new upper limit can, for the 
first time, conclusively rule out the artificial cross section enhancement of a 
factor of a hundred over the SM used in one XRB model 
\cite{2004ApJ...608L..61F}. 
This can, in turn, rule out the influence of the $^{30}$S($\alpha$,\,p) 
reaction in explaining such double-peaked burst morphology, consistent with the 
theoretical findings of a recent study \cite{2016ApJ...830...55C}.

From a technical perspective, we developed the highest quality $^{30}$S 
radioactive ion beam for astrophysical studies yet in the world.
Our analysis also showed that active target systems must be designed with 
extremely high precision capabilities and that reports of $\alpha$ scattering 
with such systems must be viewed with scrutiny.
However, the active target system enabled us to understand the energy loss 
properties of the beam very clearly, which is often a challenge for experiments 
performed using a thick target in inverse kinematics.

Further work is required to elucidate the behavior of the 
$^{30}$S($\alpha$,\,p) stellar reaction rate over the energy ranges applicable 
to XRBs so that its predicted influence on the energy generation, compositional 
inertia, and burst light curve can be reliably extracted from theoretical 
models.
Of course, a direct measurement of the $^{30}$S($\alpha$,\,p) reaction at the 
relevant energies would be the ideal approach, but it is unclear when a 
sufficiently intense, low energy $^{30}$S RIB will become available.
An intensity like $10^{4}$ pps is insufficient, and it took us four years to 
develop such an RIB for the present study.
Instead, the community should continue to exploit indirect methods as in the 
present study in the near future.
In particular, it is critical to obtain more experimental knowledge of the 
quantum properties, particularly $\Gamma_{\alpha}$, of states in $^{34}$Ar over 
$E_{\rm ex}$ from 8.0 to 11.5~MeV.

\section*{Acknowledgements}
This experiment was performed at RI Beam Factory operated by
RIKEN Nishina Center and CNS, University of Tokyo.
We appreciate the professional operation of the AVF cyclotron and the ion 
source by the RIKEN and CNS staff which made this work possible.
This work was partly supported by the Natural Sciences and Engineering Research 
Council of Canada (NSERC), National Natural Science Foundation of China (Grants 
No. 11135005 and No. 11021504), the Major State Basic Research Development 
Program of China (2013CB834406), JSPS KAKENHI (No. 21340053 and No. 16K05369) 
and the Grant-in-Aid for the Global COE Program ``The Next Generation of 
Physics, Spun from Universality and Emergence" from the Ministry of Education, 
Culture, Sports, Science and Technology (MEXT) of Japan, and the UK Science and 
Technology Facilities Council (STFC).
A.A.C. was supported in part by an Ontario Premier's Research Excellence Award 
(PREA) and by the DFG cluster of excellence ``Origin and Structure of the 
Universe" (www.universe-cluster.de).

\end{document}